\documentclass[fleqn,usenatbib]{mnras}

\pdfoutput=1

\usepackage{txfonts}

\usepackage[T1]{fontenc}
\usepackage{ae,aecompl}

\usepackage{graphicx}	
\usepackage[usenames,dvipsnames]{color}

\newcommand{\Sref}[1]{Section~\ref{#1}}
\newcommand{\Tref}[1]{Table~\ref{#1}}
\newcommand{\Aref}[1]{Appendix~\ref{#1}}
\sfcode`\.=1001\sfcode`\?=1001\sfcode`\!=1001
\newcommand{\Fref}[1]{\ifhmode \ifnum\spacefactor=1001 Figure~\ref{#1}\else Fig.~\ref{#1}\fi \else Figure~\ref{#1}\fi}
\newcommand{\Eref}[1]{\ifhmode \ifnum\spacefactor=1001 Equation~(\ref{#1})\else equation~(\ref{#1})\fi \else Equation~(\ref{#1})\fi}

\newcommand{\kms}{\ensuremath{\textrm{km\,s}^{-1}}}

\newcommand{\lya}{\ensuremath{\textrm{Ly-}\alpha}}

\newcommand{\zem}{\ensuremath{z_\textrm{\scriptsize em}}}
\newcommand{\zab}{\ensuremath{z_\textrm{\scriptsize abs}}}

\newcommand{\NHI}{\ensuremath{N_\textsc{h\scriptsize{\,i}}}}
\newcommand{\Nion}[2]{\ensuremath{N_{\rm #1\textsc{\scriptsize{\,#2}}}}}
\newcommand{\lNHI}{\ensuremath{\log(N_\textsc{h\scriptsize{\,i}}/\textrm{cm}^{-2})}}
\newcommand{\Ion}[2]{\ensuremath{\textrm{#1\,{\scshape{#2}}}}}
\newcommand{\MH}[1]{\ensuremath{\left[\textrm{#1}/\textrm{H}\right]}}
\newcommand{\ewr}[3]{\ensuremath{W_\textrm{\scriptsize r}(\Ion{#1}{#2}\,\lambda{#3})}}
\newcommand{\EWSi}{\ewr{Si}{ii}{1526}}
\newcommand{\EWC}{\ewr{C}{ii}{1334}}
\newcommand{\EBV}{\ensuremath{E(B\!-\!V)}}
\newcommand{\EBVSMC}{\ensuremath{\EBV_\textrm{\scriptsize SMC}}}
\newcommand{\EBVLMC}{\ensuremath{\EBV_\textrm{\scriptsize LMC}}}
\newcommand{\OgDLA}{\ensuremath{\Omega_\textrm{\scriptsize g}^\textrm{\scriptsize DLA}}}

\title[Dust in damped Lyman-$\alpha$ systems]{The dust content of damped Lyman-$\bmath{\alpha}$ systems in the Sloan Digital Sky Survey}

\author[M. T. Murphy \& M. L. Bernet]{Michael T. Murphy,$^{1}$\thanks{E-mail: mmurphy@swin.edu.au (MTM)} Martin L. Bernet$^{1,2}$\\
  $^{1}$Centre for Astrophysics and Supercomputing, Swinburne University of Technology, Hawthorn, Victoria 3122, Australia\\
  $^{2}$Physics Department, ETH Zurich, Wolfgang-Pauli-Strasse 27, CH-8093 Zurich, Switzerland
}

\date{Accepted 2015 October 16. Received 2015 October 15; in original form 2015 September 04}

\pubyear{2015}

\volume{{\rm in press}}

\begin{document}
\label{firstpage}
\pagerange{\pageref{firstpage}--\pageref{lastpage}}
\maketitle

\begin{abstract}
  The dust-content of damped Lyman-$\alpha$ systems (DLAs) is an
  important observable for understanding their origin and the
  neutral gas reservoirs of galaxies. While the average colour-excess
  of DLAs, \EBV, is known to be $\la$15\,milli-magnitudes (mmag), both
  detections and non-detections with $\sim$2\,mmag precision have been
  reported. Here we find 3.2-$\sigma$ statistical evidence for DLA
  dust-reddening of 774 Sloan Digital Sky Survey (SDSS) quasars by
  comparing their fitted spectral slopes to those of $\sim$7000
  control quasars. The corresponding \EBV\ is $3.0\pm1.0$\,mmag,
  assuming a Small Magellanic Cloud (SMC) dust extinction law, and it
  correlates strongly (3.5-$\sigma$) with the metal content,
  characterised by the \Ion{Si}{ii}\,$\lambda$1526 absorption-line
  equivalent width, providing additional confidence that the detection
  is due to dust in the DLAs. Evolution of \EBV\ over the redshift
  range $2.1<z<4.0$ is limited to $<$2.5\,mmag per unit redshift
  (1-$\sigma$), consistent with the known, mild DLA metallicity
  evolution. There is also no apparent relationship with neutral
  hydrogen column density, \NHI, though the data are consistent with a
  mean $\EBV/\NHI=(3.5\pm1.0)\times10^{-24}$\,mag\,cm$^{2}$,
  approximately the ratio expected from the SMC scaled to the lower
  metallicities typical of DLAs. We implement the SDSS selection
  algorithm in a portable code to assess the potential for systematic,
  redshift-dependent biases stemming from its magnitude and
  colour-selection criteria. The effect on the mean \EBV\ is
  negligible ($<$5\,per cent) over the entire redshift range of
  interest. Given the broad potential usefulness of this
  implementation, we make it publicly available.
\end{abstract}

\begin{keywords}
dust, extinction -- galaxies: high redshift -- intergalactic medium -- galaxies: ISM -- quasars: absorption lines
\end{keywords}



\section{Introduction}\label{sec:introduction}

Currently, the only direct way to probe the neutral gas surrounding
early galaxies is to study its absorption against background
objects. Damped \lya\ systems (DLAs) -- defined as having a
neutral hydrogen column densities of $\NHI=2\times10^{20}$\,cm$^{-2}$
in absorption \citep{Wolfe:1986:249} -- provide rich information about
this gas, particularly when observed along the sight-lines to quasars,
i.e.~relatively bright, very compact objects. At such column
densities, the gas in DLAs is self-shielded against the ultraviolet
background radiation from galaxies and quasars, so it is predominantly
neutral \citep[e.g.][]{Viegas:1995:268}, enabling very precise
metallic element abundances to be measured. However, some metals
(e.g.~C, Mg, Fe, Cr) will be preferentially incorporated into dust
grains \citep{Jenkins:1987:533,Pettini:1997:536}, so a complete and
accurate understanding of the nucleosynthetic history of DLA gas
requires some knowledge of its dust content. This is also an important
factor in identifying possible $\alpha$-element enhancements in DLAs
\citep[e.g.][]{Vladilo:2002:295,Ledoux:2002:802}. Further, dust grains
are the formation sites for most molecular hydrogen in cool DLA gas
\citep{Jenkins:1997:265,Ledoux:2003:209} and provide surfaces for
photoelectric heating of DLAs
\citep{Wolfe:2003:215,Wolfe:2003:235}. Understanding the dust content
of DLAs is therefore important for determining their astrophysical
origins and significance.

DLAs are cosmologically important because they contain most of the
neutral gas at all redshifts up to at least $z=5$
\citep[e.g.][]{Wolfe:1986:503,Lanzetta:1995:435,Prochaska:2005:123,Prochaska:2009:1543,Noterdaeme:2009:1087,Crighton:2015:217}. However,
if dust in DLAs obscures a significant proportion of the background
quasars from flux-limited and/or colour-selected quasar surveys
\citep{Ostriker:1984:1}, the comoving mass density of DLA gas, \OgDLA,
will be underestimated \citep[e.g.][]{Fall:1993:479}. This may be a
particularly acute concern if, as one may naively expect, higher-\NHI\
DLAs contain more dust, because DLAs with $\NHI\sim10^{21}$\,cm$^{-2}$
have the largest contribution to \OgDLA\ despite their relative
paucity (e.g.~see figures 10 and 14 of \citealt{Prochaska:2005:123}
and \citealt{Noterdaeme:2009:1087}, respectively). Surveys of DLAs in
radio-selected quasar samples avoid dust bias, though obtaining large
enough samples is much more difficult. For example, the combined
samples of \citet{Ellison:2005:1345} and \citet{Jorgenson:2006:730}
contain 26 DLAs identified towards 119 quasars and yield a \OgDLA\
which is a factor of $\sim$1.3 higher than optically-selected surveys,
but with a 33\,per cent uncertainty. Therefore, attempts to determine
\OgDLA\ with $\la$20\,per cent accuracy must incorporate other,
possibly tighter constraints on DLA dust content.

The most direct indicators of dust in DLAs would be the carbonaceous
2175\,\AA\ `dust bump' absorption feature and the silicate dust
features at 10 and 18\,$\umu$m \citep[e.g.][]{Draine:2003:241}. The
dust bump is ubiquitous in Milky Way studies but less prominent
in sight-lines through the Large Magellanic Cloud (LMC) and seemingly
absent in the Small Magellanic Cloud (SMC). Given the broad
undulations and small variations in quasar continua, clear dust bump
detections are difficult and rare. Indeed, just three detections in bona
fide DLAs have been reported
\citep{Junkkarinen:2004:658,Noterdaeme:2009:765,Ma:2015:1751}, though several have
been reported in strong \Ion{Mg}{ii} absorbers
\citep[e.g.][]{Wang:2004:589,Zhou:2010:742,Jiang:2010:1325,Wang:2012:42}. The
silicate features are also rare: the 10\,$\umu$m feature has been
detected in two bona fide DLAs
\citep{Kulkarni:2007:L81,Kulkarni:2011:14} and 5 other strong quasar
absorbers \citep{Kulkarni:2011:14,Aller:2012:19,Aller:2014:36}. Using
these features to determine the average DLA dust content appears
unlikely, if not impossible, though they clearly allow the chemical
and physical characteristics of the dust in some DLAs to be studied.

The average DLA dust content may be better determined by measuring the
reddening it should impart to the spectra of the background
quasars. The significant variation in quasar colours means that this
is an inherently statistical approach, with large samples of quasars
with foreground DLAs (`DLA quasars') and, importantly, larger samples
without DLAs (`non-DLA quasars') to act as `controls', required to
detect small amounts of dust. Early attempts
\citep{Fall:1989:7,Fall:1989:L5} culminated in a $>$4-$\sigma$
detection of average DLA reddening of $\Delta\beta=0.38\pm0.13$
\citep{Pei:1991:6} with 26 DLA quasars and 40 non-DLA quasars, where
$\beta$ represents the spectral index when the quasar continuum flux
density is fitted as
\begin{equation}\label{eq:flambda}
f_\lambda \propto \lambda^\beta\,.
\end{equation}
This implied that 10--70\,per cent of bright quasars at redshift $z=3$
were being missed in optical surveys \citep{Fall:1993:479}. The Sloan
Digital Sky Survey (SDSS) increased the available DLA and non-DLA
quasar samples significantly and ensured a more uniform quasar
selection. Using 70 DLA quasars and $\sim$1400 non-DLA quasars,
\citet{Murphy:2004:L31} found no evidence for DLA dust reddening at
$2.1<z<3.9$, with $\Delta\beta$ constrained to be $<$0.19 at
3-$\sigma$ confidence, inconsistent with previous results. Assuming an
SMC dust reddening law, the rest-frame colour excess of DLA dust
reddening was limited to $\EBVSMC < 20$\,milli-magnitudes (mmag) at
3-$\sigma$ confidence. This stringent limit indicates a very low
average DLA dust content and illustrates the large, careful analysis
required to detect dust via this statistical approach.

More recent SDSS-based analyses have provided a mix of apparent
detections and other limits on DLA dust reddening at similar redshifts
($2.2\la z\la3.5$). Using the photometric colours of 248 DLA quasars
from the SDSS Data Release (DR) 5, \citet{Vladilo:2008:701} found a
$\approx$3.5-$\sigma$ indication of reddening with rest-frame
$\EBVSMC\approx6$\,mmag. \citet{Frank:2010:2235} compared composite
spectra, made from stacking DR7 spectra in the rest-frame of 676 DLAs,
with composites from non-DLA quasars `matched' in redshift and
magnitude. This indicated no DLA reddening,
$\EBVSMC=-1.7\pm2.2$\,mmag, $\approx$2.9\,$\sigma$ lower than
\citet{Vladilo:2008:701}'s result. \citet{Khare:2012:1028} performed a
similar analysis using 1084 $\lNHI\ge20.0$ absorbers [i.e.~including
`sub-DLAs', or `super Lyman-limit systems', with $20.0\le\lNHI<20.3$],
finding no evidence for reddening from the absorbers, consistent with
\citeauthor{Frank:2010:2235}. Most recently, \citet{Fukugita:2015:195}
used the photometric colours of 1211 sub-DLA and DLA quasars from DR9
in the restricted absorption redshift range $2.1<\zab<2.3$, finding a
relatively large $\EBVSMC\approx10$\,mmag with $\sim$4-$\sigma$
significance. These recent results are discussed further in
\Sref{sec:discussion}.

Given the variety of recent results, it is important to further
consider the average dust content of DLAs. Here we use DR7 spectra to
measure the spectral index, $\beta$ in \Eref{eq:flambda}, and hence
\EBV\ for an assumed dust extinction law, in 774 SDSS DLA quasars. We
also constrain how the dust content depends on redshift, gas column
density and metal-line strength with the aim of finding clues to the
origin and physical/chemical nature of the dust and the DLAs
themselves. However, as with previous SDSS-based studies, we must
recall that the SDSS quasars targetted for spectroscopy were selected
based on their colours. In principle, this could lead to a bias
towards or against dusty DLAs (i.e.~redder spectra) in the SDSS quasar
sample. Given the complicated, redshift-dependent nature of the
colour-selection algorithm \citep{Richards:2002:2945}, this bias
should also be redshift-dependent. Here we consider in detail the
effect of colour selection bias on the measured colour excess from DLA
dust for the first time. We demonstrate the effect to be negligible at
all redshifts we consider ($2.1<\zab<4.0$).

This paper is organised as follows. \Sref{sec:samples} describes the
quasar and DLA samples used in this work and how control samples were
established for all DLA quasars. \Sref{sec:analysis} explains how the
spectral index and colour excess were measured for each
DLA. \Sref{sec:results} details the main results while discussion and
comparison with other recent results is deferred to
\Sref{sec:discussion}. We test the effects of colour-selection bias in
\Sref{sec:bias} and \Aref{app:A} describes our emulation of the SDSS
quasar colour selection algorithm of \citet{Richards:2002:2945} and
basic tests to ensure its reliability for this work. Given the broad
potential usefulness of this code, we have provided it in
\citet{Bernet:2015}. Our main conclusions are summarized in
\Sref{sec:conclusions}.

\section{Quasar and DLA sample definitions}\label{sec:samples}

\subsection{Quasar catalogues, spectra and exclusions}\label{ssec:qsos}

The results presented in this paper are derived from the SDSS DR7
versions of the spectra identified as quasars in the catalogue of
\citet{Schneider:2010:2360}. This catalogue contains 105783 quasars
with improved (where necessary) emission redshift measurements
compared with those automatically determined by the original SDSS data
processing system. We exclude the 6214 broad absorption line (BAL)
quasars identified in this catalogue by \citet{Shen:2011:45} because
the BAL features may have interfered with the DLA identification
procedure (see \Sref{ssec:dlas}) and because BAL quasars are often
heavily reddened compared to the normal distribution of quasar
colours \citep[e.g.][]{Reichard:2003:1711}. We also exclude quasars
with emission redshifts outside the range $2.20\le\zem\le4.42$. The
lower redshift cut ensures a non-zero search path for DLAs redwards of
the blue limit of the SDSS spectra (i.e.~$\approx$3800\,\AA) while the
higher redshift cut ensures that the reddest band we use to determine
the quasar spectral index (quasar rest-frame wavelengths
1684--1700\,\AA; see \Sref{sec:analysis}) remains entirely bluewards
of the red limit of the SDSS spectra (i.e.~$\sim$9200\,\AA). These
redshift restrictions leave 14672 quasars (note the further restrictions
on the quasar sample below in \Sref{ssec:dlas}).

We also cross-checked our main results using both the DR5 and DR7
spectra of the quasars identified in the DR5 quasar catalogue of
\citet{Schneider:2007:102}, and find consistent results. A new
reduction and spectrophotometric calibration was introduced in DR6,
yielding a 30\,mmag root-mean-square (RMS) difference between the
$r-i$ colours from the point-spread-function (PSF) photometry and
those synthesized from the calibrated spectra of stars\footnote{See
  description at
  {\urlstyle{rm}\url{http://www.sdss2.org/dr7/algorithms/spectrophotometry.html}}.}. The
spectrophotometry of DR9 spectra is expected to be biased because
smaller fibres were used (2 vs.~3\arcsec\ diameter, respectively) and,
without atmospheric dispersion compensation available, they were
offset to collect blue light at the expense of red light to probe the
\lya\ forest \citep{Paris:2012:66}.

\subsection{DLA samples}\label{ssec:dlas}

The baseline results in this paper were derived using DLAs from the
catalogue of \citet{Noterdaeme:2009:1087}. They searched DR7 spectra
of objects that the automatic SDSS data processing labelled as quasars
with $\zem>2.17$. The search was restricted to 8339 spectra with
median continuum-to-noise ratios (CNRs) exceeding 4\,per pixel
(pix$^{-1}$) redwards of the wavelength where the signal-to-noise
ratio (SNR) first exceeds 4\,pix$^{-1}$. \citet{Noterdaeme:2009:1087}
applied an automated algorithm to detect DLAs and determine their
\zab, \NHI\ and rest equivalent widths of selected metal lines,
e.g.~\EWC\ and more often \EWSi.

From the 14672 DR7 quasars selected in \Sref{ssec:qsos} we removed the
6743 quasars that were not among those searched for `strong absorbers'
by \citet{Noterdaeme:2009:1087}, leaving 7929 quasars. These `strong
absorbers' include DLAs and sub-DLAs systems with
$\lNHI\ge20.0$\,cm$^{-2}$. Our main results and much of our discussion
below focus on the DLAs only. However, as a cross-check on possible
systematic effects and to boost sample sizes when searching for trends
within our results, we also perform the same analyses with the
sub-DLAs included.

The final restriction on the quasar and absorber samples is to remove
quasar spectra in which \citet{Noterdaeme:2009:1087} identified more
than one `strong absorber'. This last selection criterion defines our
baseline samples as follows:
\begin{itemize}
\item \textbf{DLA \& non-DLA samples:} 7879 quasars;
  774 are DLA quasars (i.e.~with a single, bona fide
  DLA); 7105 are non-DLA quasars.
\item \textbf{Absorber \& non-absorber samples:} 7802 quasars; 1069
  are absorber quasars [i.e.~with a single
  $\lNHI\ge20.0$\,cm$^{-2}$ system]; 6733 are non-absorber quasars.
\end{itemize}
Note that this last criterion means that the DLA$+$non-DLA sample is
larger than the absorber$+$non-absorber sample. This is because
sub-DLAs are more common than DLAs, and because sub-DLAs are allowed
in the DLA and non-DLA quasar spectra but only a single DLA or sub-DLA
is allowed in the absorber quasars. That DLA quasars are allowed to be
`contaminated' by sub-DLAs is offset by the contamination of non-DLA
quasars in the same way and in approximately equal proportion, thereby
leaving a negligible residual effect on the differential reddening
measurement we make in \Sref{sec:analysis}.

\begin{figure}
\begin{center}
\includegraphics[width=1.0\columnwidth]{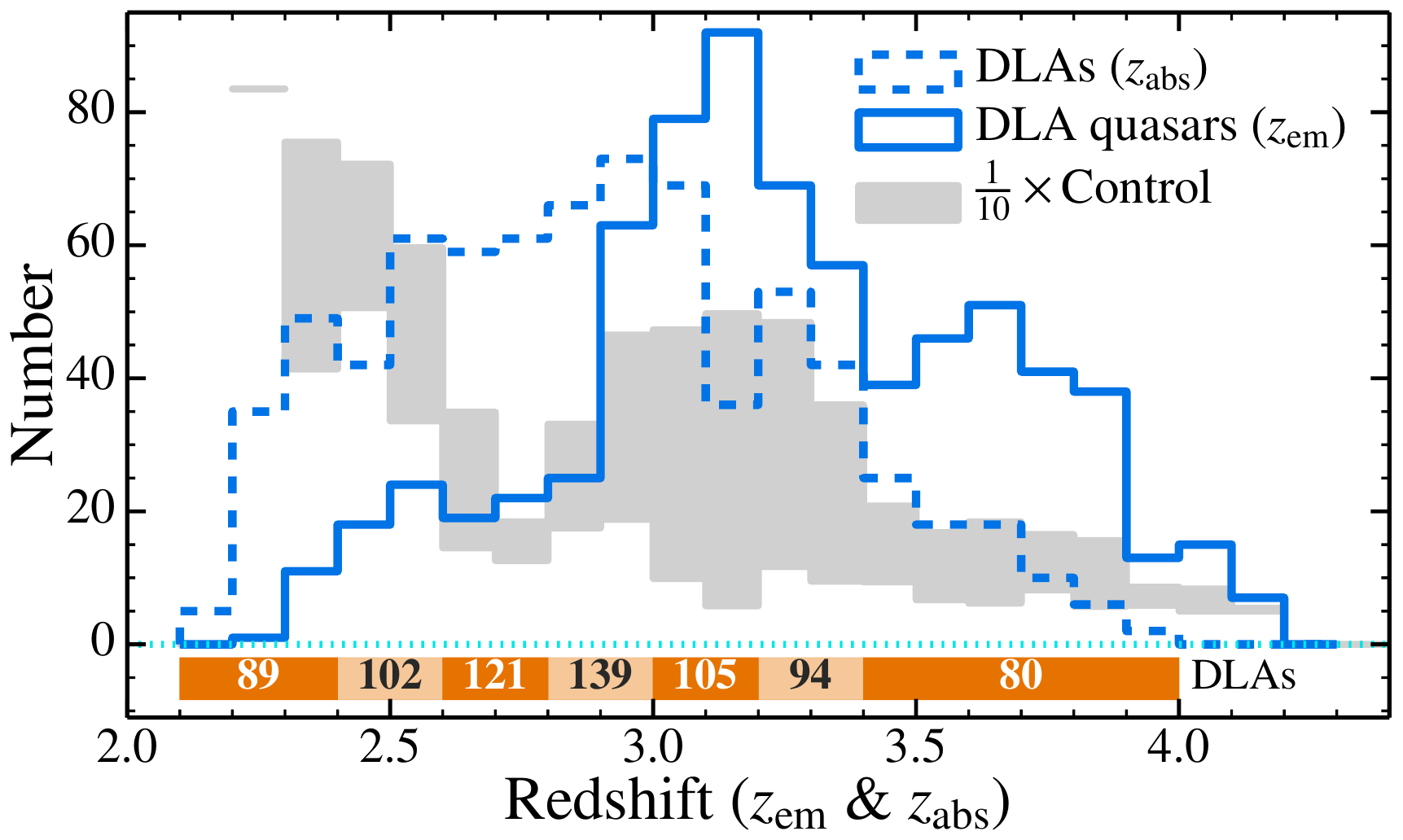}\vspace{-0.5em}
\caption{Emission redshift (\zem) distributions of DLA quasars (solid
  line) and their corresponding control quasars (shaded regions), and
  the absorption redshift (\zab) distribution of the DLAs
  themselves. There is a single DLA per DLA quasar. For each $\Delta
  z=0.1$ bin, the shaded region indicates the range in sizes of the
  control samples for the DLA quasars in that bin. Note that these
  shaded regions are normalized by a factor of 10 for clarity; for
  example, the minimum and maximum number of control quasars
  corresponding to any of the DLA quasars with $3.0\le\zem<3.1$ is 99
  and 473, respectively. The shaded bar below the histograms, labelled
  with the number of DLAs in each bin, shows the absorption redshift
  bins used in subsequent figures.}
\label{fig:redshift_dist}
\end{center}
\end{figure}

Tables \ref{tab:DLA_sample}--\ref{tab:Non-Abs_sample} provide the
relevant information for each of these samples, including the basic
(sub-)DLA parameters used in the reddening analysis below.
\Fref{fig:redshift_dist} shows the redshift distributions of the DLA
quasars and the DLAs themselves. Note the relative paucity of DLA
quasars with $\zem<2.9$: quasar colours cross the stellar locus
between $\zem\sim2.5$--3.0, so the SDSS colour-selection algorithm had
to be specially modified there even to select some quasars. This is
one example of the strongly varying, complicated quasar colour
selection function in the SDSS. By comparison,
\Fref{fig:redshift_dist} shows that the DLA redshift distribution is
relatively smooth.

\newcommand\oldtabcolsep{\tabcolsep}
\setlength{\tabcolsep}{0.35em}

\begin{table*}
\begin{center}
\caption{The DLA quasar sample (774 quasars). The quasars are listed
  by their SDSS name (formed from right ascension and declination),
  emission redshift (\zem), the redshift search path for DLAs
  defined by $z_{\rm min}$ and $z_{\rm max}$ [see \Eref{eq:zcrit2}],
  the measured spectral index, $\beta$, and its 1-$\sigma$
  uncertainty ($\sigma_\beta$), and the spectral index measured with
  an unweighted power-law fit, $\beta_{\rm nw}$. The DLA properties
  required in our analysis are the absorption redshift (\zab),
  logarithmic column density,
  $\mathcal{N}_\textsc{h\scriptsize{\,i}}\equiv\lNHI$, the
  rest-frame equivalent widths \EWSi\ and \EWC\ (with non-detections
  indicated by ``$-999.00$''), the measured difference, $\Delta\beta$,
  between the DLA quasar's $\beta$ and the mean $\beta$ of its
  control distribution, the corresponding colour excess, \EBV, for
  SMC and LMC-like dust, and the number of control quasars ($N_{\rm
    ctrl}$). The full table is available in the electronic version
  of this paper (see the Supporting Information).}
\label{tab:DLA_sample}\vspace{-0.5em}
\begin{tabular}{lcccccccccccccc}\hline
\multicolumn{1}{c}{SDSS name}&
\multicolumn{1}{c}{\zem}&
\multicolumn{1}{c}{$z_{\rm min}$}&
\multicolumn{1}{c}{$z_{\rm max}$}&
\multicolumn{1}{c}{$\beta$}&
\multicolumn{1}{c}{$\sigma_\beta$}&
\multicolumn{1}{c}{$\beta_{\rm nw}$}&
\multicolumn{1}{c}{\zab}&
\multicolumn{1}{c}{$\mathcal{N}_\textsc{h\scriptsize{\,i}}$}&
\multicolumn{2}{c}{$W_{\rm r}$ [\AA]}&
\multicolumn{1}{c}{$\Delta\beta$}&
\multicolumn{2}{c}{\EBV\ [mag]}&
\multicolumn{1}{c}{$N_{\rm ctrl}$}\\
\multicolumn{1}{c}{(J2000)}& & & & & & & &
&
\multicolumn{1}{c}{\Ion{Si}{ii}\,$\lambda$1526}&
\multicolumn{1}{c}{\Ion{C}{ii}\,$\lambda$1334}&
&
\multicolumn{1}{c}{SMC}&
\multicolumn{1}{c}{LMC}&
\\
\hline
001240.57$+$135236.7 & $3.1866$ & $2.371$ & $3.116$ & $-1.266$ & $0.104$ & $-1.273$ & $3.022$ & $20.86$ & $-999.00$ & $-999.00$ & $ 0.195$ & $ 0.0130$ & $ 0.0184$ & 497\\
001328.20$+$135828.0 & $3.5755$ & $2.501$ & $3.499$ & $-1.371$ & $0.060$ & $-1.389$ & $3.281$ & $21.56$ & $   0.30$ & $   0.34$ & $ 0.073$ & $ 0.0051$ & $ 0.0078$ & 160\\
001813.89$+$142455.6 & $4.2355$ & $3.409$ & $4.147$ & $-1.672$ & $0.170$ & $-1.184$ & $3.861$ & $20.65$ & $-999.00$ & $-999.00$ & $-0.162$ & $-0.0110$ & $-0.0166$ &  29\\
003501.88$-$091817.6 & $2.4187$ & $2.224$ & $2.361$ & $-1.608$ & $0.080$ & $-1.601$ & $2.337$ & $20.42$ & $-999.00$ & $-999.00$ & $-0.063$ & $-0.0041$ & $-0.0059$ & 545\\
005319.20$+$134708.8 & $2.9203$ & $2.275$ & $2.855$ & $-1.332$ & $0.077$ & $-1.332$ & $2.628$ & $20.92$ & $-999.00$ & $   0.62$ & $ 0.146$ & $ 0.0107$ & $ 0.0175$ & 352\\
\hline
\end{tabular}
\end{center}
\end{table*}

\begin{table*}
\begin{center}
\caption{The absorber quasar sample (1069 quasars). The columns are
  defined in the caption of \Tref{tab:DLA_sample} and the full table
  is available in the electronic version of this paper (see the
  Supporting Information).}
\label{tab:Abs_sample}\vspace{-0.5em}
\begin{tabular}{lcccccccccccccc}\hline
\multicolumn{1}{c}{SDSS name}&
\multicolumn{1}{c}{\zem}&
\multicolumn{1}{c}{$z_{\rm min}$}&
\multicolumn{1}{c}{$z_{\rm max}$}&
\multicolumn{1}{c}{$\beta$}&
\multicolumn{1}{c}{$\sigma_\beta$}&
\multicolumn{1}{c}{$\beta_{\rm nw}$}&
\multicolumn{1}{c}{\zab}&
\multicolumn{1}{c}{$\mathcal{N}_\textsc{h\scriptsize{\,i}}$}&
\multicolumn{2}{c}{$W_{\rm r}$ [\AA]}&
\multicolumn{1}{c}{$\Delta\beta$}&
\multicolumn{2}{c}{\EBV\ [mag]}&
\multicolumn{1}{c}{$N_{\rm ctrl}$}\\
\multicolumn{1}{c}{(J2000)}& & & & & & & &
&
\multicolumn{1}{c}{\Ion{Si}{ii}\,$\lambda$1526}&
\multicolumn{1}{c}{\Ion{C}{ii}\,$\lambda$1334}&
&
\multicolumn{1}{c}{SMC}&
\multicolumn{1}{c}{LMC}&
\\
\hline
001240.57$+$135236.7 & $3.1866$ & $2.371$ & $3.116$ & $-1.266$ & $0.104$ & $-1.273$ & $3.022$ & $20.86$ & $-999.00$ & $-999.00$ & $ 0.196$ & $ 0.0125$ & $ 0.0176$ & 457\\
001328.20$+$135828.0 & $3.5755$ & $2.501$ & $3.499$ & $-1.371$ & $0.060$ & $-1.389$ & $3.281$ & $21.56$ & $   0.30$ & $   0.34$ & $ 0.091$ & $ 0.0062$ & $ 0.0092$ & 140\\
001813.89$+$142455.6 & $4.2355$ & $3.409$ & $4.147$ & $-1.672$ & $0.170$ & $-1.184$ & $3.861$ & $20.65$ & $-999.00$ & $-999.00$ & $-0.214$ & $-0.0162$ & $-0.0261$ &  26\\
003126.79$+$150739.5 & $4.2832$ & $3.432$ & $4.195$ & $-2.904$ & $0.156$ & $-2.427$ & $4.083$ & $20.28$ & $-999.00$ & $-999.00$ & $-1.512$ & $-0.0984$ & $-0.1367$ &  31\\
003501.88$-$091817.6 & $2.4187$ & $2.224$ & $2.361$ & $-1.608$ & $0.080$ & $-1.601$ & $2.337$ & $20.42$ & $-999.00$ & $-999.00$ & $-0.063$ & $-0.0043$ & $-0.0062$ & 538\\
\hline
\end{tabular}
\end{center}
\end{table*}

\begin{table}
\begin{center}
\caption{The non-DLA quasar sample (7105 quasars). The columns are
  defined in the caption of \Tref{tab:DLA_sample} and the full table
  is available in the electronic version of this paper (see the
  Supporting Information).}
\label{tab:Non-DLA_sample}\vspace{-0.5em}
\begin{tabular}{lcccccc}\hline
\multicolumn{1}{c}{SDSS name (J2000)}&
\multicolumn{1}{c}{\zem}&
\multicolumn{1}{c}{$z_{\rm min}$}&
\multicolumn{1}{c}{$z_{\rm max}$}&
\multicolumn{1}{c}{$\beta$}&
\multicolumn{1}{c}{$\sigma_\beta$}&
\multicolumn{1}{c}{$\beta_{\rm nw}$}\\
\hline
000050.60$-$102155.9 & $2.6404$ & $2.136$ & $2.579$ & $-2.056$ & $0.045$ & $-2.018$\\
000143.41$+$152021.4 & $2.6383$ & $2.152$ & $2.578$ & $-1.150$ & $0.069$ & $-1.158$\\
000221.11$+$002149.3 & $3.0699$ & $2.136$ & $3.001$ & $-1.422$ & $0.053$ & $-1.428$\\
000300.34$+$160027.6 & $3.6983$ & $3.545$ & $3.619$ & $-1.496$ & $0.145$ & $-1.088$\\
000413.64$-$085529.6 & $2.4241$ & $2.153$ & $2.366$ & $-1.300$ & $0.037$ & $-1.323$\\
\hline
\end{tabular}
\end{center}
\end{table}

\begin{table}
\begin{center}
\caption{The non-absorber quasar sample (6733 quasars). The columns are
  defined in the caption of \Tref{tab:DLA_sample} and the full table
  is available in the electronic version of this paper (see the
  Supporting Information).}
\label{tab:Non-Abs_sample}\vspace{-0.5em}
\begin{tabular}{lcccccc}\hline
\multicolumn{1}{c}{SDSS name (J2000)}&
\multicolumn{1}{c}{\zem}&
\multicolumn{1}{c}{$z_{\rm min}$}&
\multicolumn{1}{c}{$z_{\rm max}$}&
\multicolumn{1}{c}{$\beta$}&
\multicolumn{1}{c}{$\sigma_\beta$}&
\multicolumn{1}{c}{$\beta_{\rm nw}$}\\
\hline
000050.60$-$102155.9 & $2.6404$ & $2.136$ & $2.579$ & $-2.056$ & $0.045$ & $-2.018$\\
000143.41$+$152021.4 & $2.6383$ & $2.152$ & $2.578$ & $-1.150$ & $0.069$ & $-1.158$\\
000221.11$+$002149.3 & $3.0699$ & $2.136$ & $3.001$ & $-1.422$ & $0.053$ & $-1.428$\\
000300.34$+$160027.6 & $3.6983$ & $3.545$ & $3.619$ & $-1.496$ & $0.145$ & $-1.088$\\
000413.64$-$085529.6 & $2.4241$ & $2.153$ & $2.366$ & $-1.300$ & $0.037$ & $-1.323$\\
\hline
\end{tabular}
\end{center}
\end{table}

\setlength{\tabcolsep}{\oldtabcolsep}

While the SDSS spectral resolving power ($R\approx1800$) and the CNR
and SNR thresholds ensure that most DLAs are identified, they are not
optimal for accurately estimating \NHI. To cross-check our main
results against different approaches to estimating \NHI\ we used the
alternative, albeit smaller sample of DLAs from DR5 by
\citet{Prochaska:2009:1543}; it provides results consistent with those
presented here. Although their DLA search employed similar SNR
threshold criteria to \citet{Noterdaeme:2009:1087}, their approach to
fitting the DLAs to estimate \NHI\ differs, so we expect that a
comparison of results from the two samples is a reasonable test for
biases, particularly at low \NHI. Using the `statistical sample' from
\citet{Prochaska:2009:1543}, and a similar series of exclusions as we
applied above to the \citet{Noterdaeme:2009:1087} sample, left 526
DLA quasars and 7712 non-DLA quasars.

\subsection{Control samples}\label{ssec:controls}

From the non-DLA samples above (\Sref{ssec:dlas}), a sample of
``control'' quasars is selected for \textit{each} DLA quasar to
benchmark the typical distribution of quasar spectral indices expected
in the absence of strong absorbers. To be selected as a control quasar
for a given DLA quasar with redshift $\zem^{\rm DLA}$ and DLA redshift
$\zab^{\rm DLA}$, a non-DLA quasar with redshift \zem\ must satisfy
the following criteria:
\begin{equation}\label{eq:zcrit1}
\zem^{\rm DLA}-0.05 \le \zem \le \zem^{\rm DLA}+0.05;~~{\rm and}
\end{equation}\vspace{-2em}
\begin{equation}\label{eq:zcrit2}
z_{\rm min} \le \zab^{\rm DLA} \le z_{\rm max}\,.
\end{equation}
Here, $z_{\rm min}$ and $z_{\rm max}$ define, respectively, the
minimum and maximum redshifts between which the non-DLA quasar
spectrum was searched for DLAs by \citet{Noterdaeme:2009:1087}. These
values were kindly provided to us by P.~Noterdaeme. The maximum
redshift, $z_{\rm max}$, was defined for all quasars to be 5000\,\kms\
bluewards of \zem\ so as to avoid strong absorbers associated with the
quasars themselves. This means that both the DLA and non-DLA samples
defined above may have such ``associated'' DLAs and sub-DLAs in their
spectra. However, again, this will occur with very similar frequency
in both the DLA and non-DLA samples, so the differential reddening
measurement will not be significantly affected. Similar criteria are
applied when forming the control samples for absorber quasars from the
non-absorber sample. The minimum redshift, $z_{\rm min}$, was defined
by \citeauthor{Noterdaeme:2009:1087} according to the CNR and SNR
criteria mentioned in \Sref{ssec:dlas}. This means that $z_{\rm min}$
effectively acts as a spectral quality control measure to ensure that
DLAs can be securely identified between $z_{\rm min}$ and $z_{\rm
  max}$.

The first criterion above reflects the requirement for control quasars
to have similar redshifts as their corresponding DLA quasar. It
defines a small redshift interval within which we assume the
complicated, redshift-dependent SDSS quasar selection biases are
relatively constant. It effectively determines the number of control
quasars for a given DLA quasar; increasing it increases the control
sample size but risks larger effects from non-uniformities in the
quasar selection. Reducing it substantially below $\pm$0.05 decreases
the control samples of many DLA quasars. We find that our results are
insensitive to changes in this interval up to $\pm$0.2.

The second criterion above states that a non-DLA quasar can only
qualify as a control quasar for a given DLA quasar (with a DLA at
\zab) if a hypothetical DLA at \zab\ could have been found within in
its spectrum. For that to be true, \zab\ must lie between the $z_{\rm
  min}$ and $z_{\rm max}$ values of the non-DLA quasar. Inherent in
this criterion is the assumption that DLA dust extinction and
reddening are small: if they were large, and such a dusty DLA was
placed between $z_{\rm min}$ and $z_{\rm max}$ in a non-DLA spectrum,
it would suppress the continuum flux in the \lya\ forest region, and
$z_{\rm min}$ would increase, potentially to the point where the
second criterion is no longer satisfied. In this sense, the CNR and
SNR thresholds in the DLA search algorithm, combined with the second
criterion above, may slightly diminish the measured DLA dust reddening
signal compared to its true strength. However, the bias will reduce in
proportion to the reddening itself, and will likely be negligible at
the very small reddening levels observed in our work. Simple tests, in
which $z_{\rm min}$ was artificially varied for the DLA sample,
confirmed this expectation.

For the statistical quasar reddening analysis below in
\Sref{sec:analysis}, each DLA quasar was conservatively required to
have at least 50 control quasars. This excludes 54 DLA quasars and 78
absorber quasars from the baseline samples of 774 and 1069 defined
above. \Fref{fig:redshift_dist} shows the resulting range in control
sample sizes for the DLA quasars as a function of redshift. The median
number of control quasars is 264 per DLA quasar but, like the DLA
quasar distribution, this varies considerably with redshift. For
example, we again notice the strong drop in control quasar numbers
between $\zem\sim2.6$--3.0 because of the similarity between quasar
and stellar colours there. Note that many DLA quasars may share many
control quasars in common.

Finally, we note that other recent SDSS DLA reddening studies
\citep[e.g.][]{Vladilo:2008:701,Frank:2010:2235,Khare:2012:1028}
imposed an explicit magnitude restriction on control quasars: they
required that a control quasar have a similar $i$-band magnitude to
the corresponding DLA quasar magnitude. We do not impose such a
criterion in our analysis, though we emphasise that the second
criterion above ensures that control quasars have the same CNR
distribution in the Lyman-$\alpha$ forest region as the DLA quasars at
the same redshift. We discuss the implications of matching the DLA
and control quasar $i$-band magnitudes in \Sref{sec:discussion}.

\section{Statistical quasar reddening analysis}\label{sec:analysis}

\begin{figure}
\begin{center}
\includegraphics[width=1.0\columnwidth]{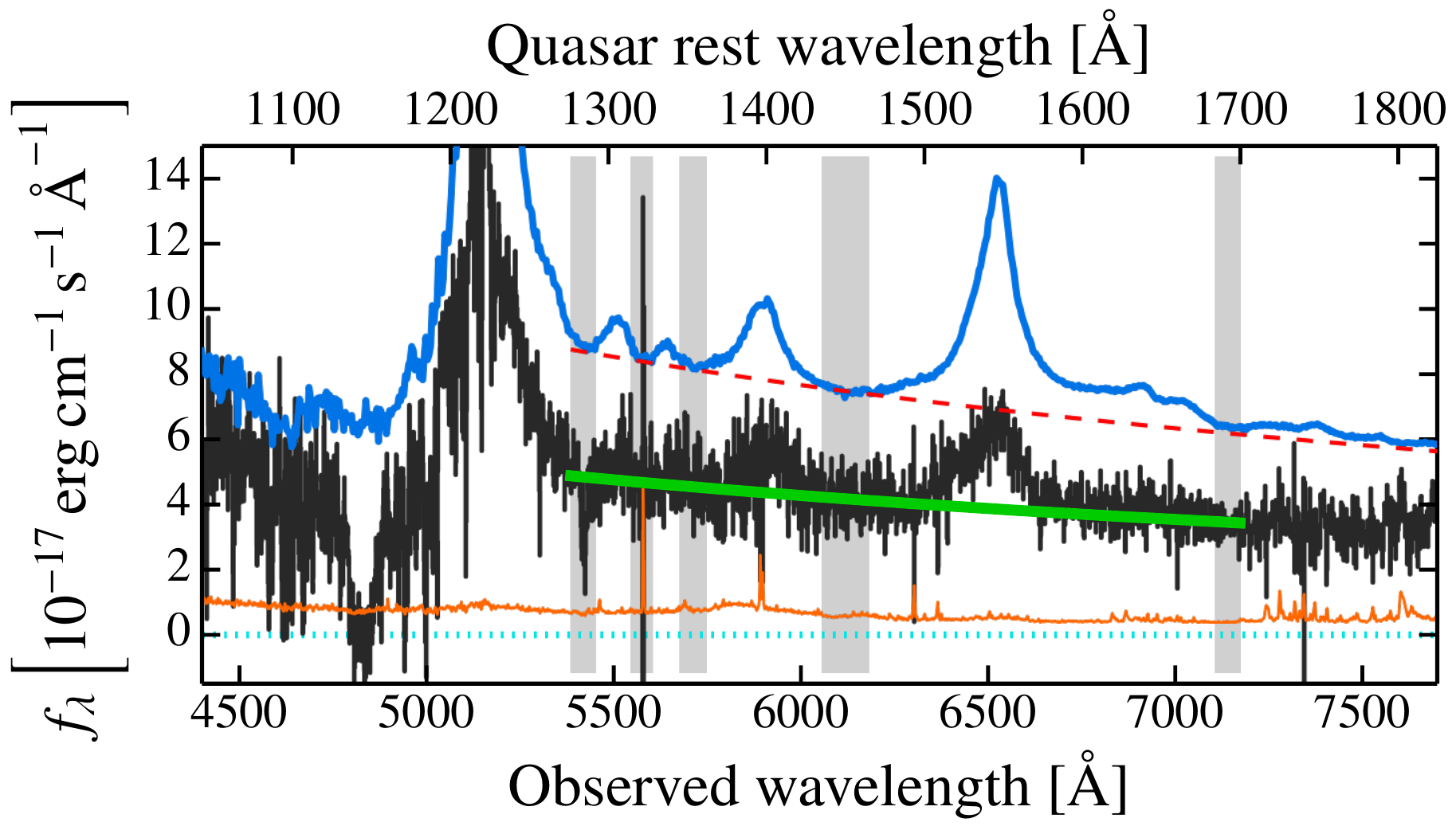}\vspace{-0.5em}
\caption{Example of spectral index fitting on the SDSS spectrum
  (black/dark solid line) of quasar J112820.20$+$325918.9
  ($\zem=3.22$) which has a DLA at $\zab=2.909$ with $\lNHI=20.6$
  \citep{Noterdaeme:2009:1087}. The power-law (green/grey thick line)
  was fitted to the median fluxes within the grey shaded regions and
  has a spectral index $\beta=-1.22$. The orange/thin lower line is
  the 1-$\sigma$ flux uncertainty. The blue/upper curve shows the
  median composite SDSS quasar spectrum of \citet{VandenBerk:2001:549}
  and their power-law estimate of the underlying continuum (dashed
  line, $\beta=-1.56$). Note that the grey shaded fitting regions are
  chosen to match areas of little or no quasar line emission.}
\label{fig:specplot}
\end{center}
\end{figure}

The spectral index, $\beta$ in \Eref{eq:flambda}, was determined for
each DLA quasar and all its control quasars following a similar
approach to \citet{Murphy:2004:L31}, as illustrated in
\Fref{fig:specplot}. Before performing the spectral fitting below, all
quasar spectra were first corrected for Galactic extinction using the
dust map of \citet{Schlegel:1998:525} and the Milky Way extinction law
of \citet{ODonnell:1994:158} \citep[based on that
of][]{Cardelli:1989:245}.

For each quasar, the median flux density was determined in 5 small
spectral regions (grey shaded regions in \Fref{fig:specplot}) at
quasar rest-frame wavelengths of 1276--1292, 1314--1328, 1345--1362,
1435--1465, 1684--1700\,\AA. As shown in \Fref{fig:specplot}, the
composite SDSS spectrum of \citet{VandenBerk:2001:549} in these
regions shows little or no evidence for contamination from quasar line
emission and is consistent with representing the underlying power-law
quasar continuum. Using the median $f_\lambda$ value in each region
provides robustness against the effects of narrow absorption features,
bad pixels, residuals from the removal of telluric features and/or
other narrow artefacts. The second region (1314--1328\,\AA) is the
narrowest, but still comprises 47 SDSS pixels (69\,\kms\,pix$^{-1}$),
so the median is substantially effective in guarding against these
effects. As a measure of the relative uncertainty in the median
fluxes, we also calculate the semi-interquartile range, i.e.~half the
range around the median containing half of the $f_\lambda$ values, in
each region. These will be small for high SNR spectra and larger for
low SNR spectra and/or regions that contain some features/artefacts
extending over more than $\sim$10 pixels.

\begin{figure}
\begin{center}
\includegraphics[width=1.0\columnwidth]{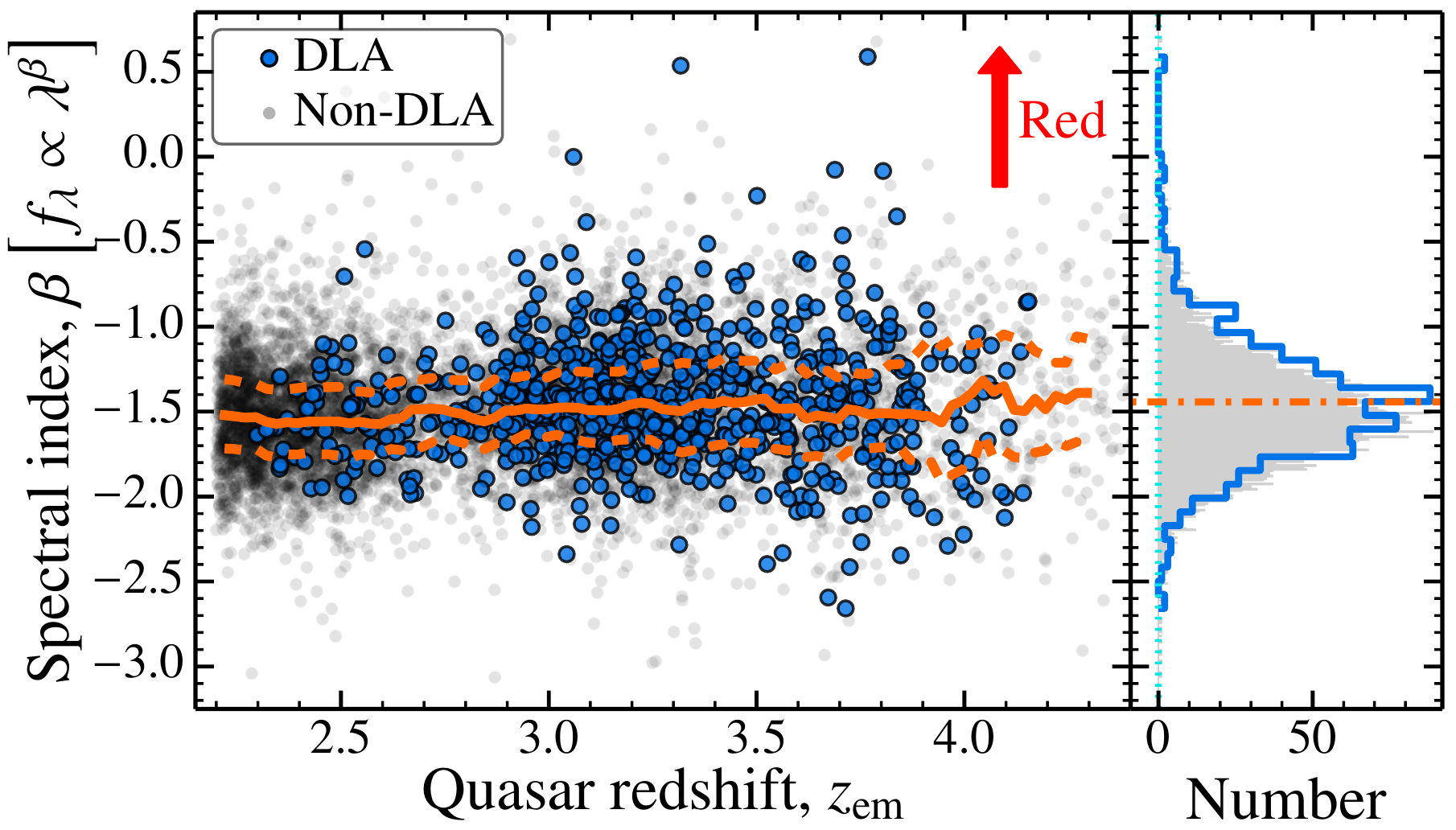}\vspace{-0.5em}
\caption{Distribution of spectral index, $\beta$, for the DLA quasars
  (dark, circled points) and non-DLA quasars (light grey points) with
  emission redshift. The DLA distribution clearly follows the shape of
  the non-DLA distribution closely. A running median (over 201
  quasars) of the non-DLA quasar $\beta$ values is plotted as an
  orange/grey solid line and shows some evolution with redshift (the
  dashed lines mark the running upper and lower quartiles). The
  overall $\beta$ distributions are shown in the histograms
  (right-hand panel). The two distributions are similar and both
  display a skew towards redder $\beta$ values. The dot-dashed line
  shows the mean $\beta$ for the DLA quasars.}
\label{fig:beta_dist}
\end{center}
\end{figure}

We estimated $\beta$ using a weighted power-law fit to these median
$f_\lambda$ values with the inverse-squares of their
semi-interquartile ranges as weights. This provides a robust,
naturally-weighted $\beta$ estimate with a representative uncertainty,
$\delta\beta$. These uncertainties have a median of 0.13, with
semi-interquartile range of 0.05 and no strong redshift
dependence. \Fref{fig:beta_dist} shows the distribution of $\beta$
with redshift for the DLA quasars and non-DLA quasars (the latter is
the pool from which control quasars are selected). Note that $\beta$
evolves little with redshift in the DR7 quasar spectra. Relative to
the $\beta$ distribution of the non-DLA quasars, there appears to be a
small shift in the DLA quasar $\beta$ distribution to redder colours
at most redshifts. This is an indication that the DLAs cause their
quasars to be reddened with respect to non-DLA quasars at similar
redshifts (whose spectra had sufficient SNR for those DLAs to be
detected). The reddening effect is substantially smaller than the
width of the $\beta$ distribution itself, highlighting the need for
large statistical samples of both DLA and non-DLA quasars.

One potential concern with our spectral fitting procedure is that the
relative weights of the different fitting regions will change with
redshift. In particular, the redder fitting regions will move into
noisier parts of the spectrum at higher redshifts, so they will
contribute less to the weighted power-law fit to determine $\beta$. In
principle, this may introduce a redshift-dependency in the spectral
index results. To check this we determined the spectral index with an
unweighted fit, $\beta_{\rm unweighted}$, and analysed the difference,
$\beta-\beta_{\rm unweighted}$, for DLA quasars and non-DLA quasars as
a function of redshift. There is no significant difference between the
weighted and unweighted $\beta$ values up to $\zem\approx3.3$, but
above this redshift $\beta$ is smaller than $\beta_{\rm unweighted}$
by $\approx$0.1 on average. However, most importantly, this behaviour
is the same for both the DLA and non-DLA samples, i.e.~there is no
significant \textit{differential} changes between the DLA quasar and
non-DLA quasar samples with redshift. It is important to realise that
variations in $\beta$ -- either real or spurious -- over redshift
intervals greater than the $\Delta z=0.1$ bin size for selecting
control quasars, will not affect the $\Delta\beta$ value for a given
DLA quasar. We conclude that the weighted approach to determining the
spectral index of each quasar spectrum is robust, at least to the
precision required for the differential reddening analysis here.

If DLAs contain negligible dust, the distribution of $\beta$ for DLA
quasars and their control quasars should be
indistinguishable. Significant DLA dust reddening should increase a
DLA quasar's spectral index, $\beta_{\rm DLA}$, relative to the
$\beta$ distribution of its control quasars. That is, we should seek
to measure
\begin{equation}\label{eq:Deltabeta}
\Delta\beta\equiv\beta_{\rm DLA}-\left<\beta_{\rm control}\right>\,
\end{equation}
for each DLA using some appropriate definition of $\left<\beta_{\rm
    control}\right>$, a statistic representing the $\beta$
distribution of that DLA quasar's control sample. \Fref{fig:beta_dist}
shows that the $\beta$ distribution of non-DLA quasars is non-Gaussian
and asymmetric. However, it also shows that the $\beta_{\rm DLA}$
distribution has a very similar shape. We therefore make the simple
assumption in our analysis that the two distributions differ only by
an average shift in the $\beta$ distribution. This implies that
$\left<\beta_{\rm control}\right>$ in \Eref{eq:Deltabeta} is just the
simple mean $\beta$ of the control sample for each DLA quasar. For
example, consider the limiting case of no DLA reddening: $\beta_{\rm
  DLA}$ in \Eref{eq:Deltabeta} is then equivalent to a single value
drawn from the $\beta_{\rm control}$ distribution; considering many
realizations of the same DLA quasar, the mean $\Delta\beta$ will be
zero only when $\left<\beta_{\rm control}\right>$ is the mean $\beta$
of the control sample. Simple Monte Carlo experiments confirm this,
using a variety of underlying $\beta$ distributions, even highly
asymmetric ones, as long as statistical DLA reddening is assumed to
simply shift the $\beta$ distribution of non-DLA quasars.

The above approach allows us to measure a value of $\Delta\beta$ for
every DLA quasar, though it must be emphasised that individual
$\Delta\beta$ measurements are not particularly meaningful and that a
statistical $\Delta\beta$ distribution is needed to infer the average
dust content of DLAs. One advantage of this approach is that some
heavily reddened DLA quasars might be identifiable in this
analysis. It also enables a simple conversion between $\Delta\beta$
and the (physically more interesting) colour excess, \EBV, using
different dust extinction laws (e.g.~SMC and LMC) for each DLA -- see
below.\footnote{An alternative approach to determining the average
  reddening as a function of some physically interesting quantity,
  e.g.~\NHI, would be to pre-define bins of \NHI\ and compare the
  distribution of $\beta_{\rm DLA}$ for DLAs in each bin with the
  collective, appropriately weighted $\beta$ distribution of the
  relevant control quasars. This could be done, for example, with a
  maximum likelihood measure of the shift in $\beta$ required to align
  the two distributions in each pre-defined bin. However, this would
  provide less flexibility and transparency in the results in
  \Sref{sec:results} compared to our approach of defining
  $\Delta\beta$ appropriately and measuring it for each DLA quasar.} A
more sophisticated definition of $\left<\beta_{\rm control}\right>$
might incorporate the $\beta$ uncertainties, $\delta\beta$, for the
individual control quasars. However, this is only likely to affect the
results if $\delta\beta$ is a strong function of $\beta$; we do not
observe such a relationship in our spectral fitting results, so we use
the simple mean for $\left<\beta_{\rm control}\right>$.

Finally, for each DLA quasar, we convert the $\Delta\beta$ measurement
from \Eref{eq:Deltabeta} into DLA rest-frame colour excess value,
\EBVSMC\ for SMC-like dust. We redden and de-redden the DLA quasar
spectrum with a wide range of \EBV\ values (both positive and
negative, respectively), applied in the rest-frame of the DLA, and
re-measure $\beta$ for each value to establish a one-to-one
relationship between \EBV\ and changes in $\beta$. Interpolating this
relationship then allows conversion of an arbitrary $\Delta\beta$
measurement into a \EBVSMC\ value for that DLA quasar. The same
approach is used to derive a colour excess value for LMC-like dust,
\EBVLMC. We use the SMC and LMC extinction laws from
\citet[][specifically the fitting formulae in their table
4]{Pei:1992:130} for this conversion. The previous spectral stacking
analyses of SDSS DLAs have shown no significant evidence for a
2175\,\AA\ dust bump \citep{Frank:2010:2235,Khare:2012:1028}, so we do
not include a Milky Way dust law in our analysis.

\section{Results}\label{sec:results}

The main numerical results of this work are summarized in
\Tref{tab:res} and detailed in the following subsections. Correlations
and linear relationships between the measured quantities --
i.e.~$\Delta\beta$ and \EBV\ using SMC and LMC dust models -- and
\zab, \NHI, metal-line equivalent widths etc.~are derived using the
individual measurements, while binned measurements are plotted in the
main figures. Because the scatter in the $\Delta\beta$ and \EBV\
measurements is completely dominated by the underlying diversity of
$\beta$ values (see \Fref{fig:beta_dist}), the linear relationships
are derived using unweighted linear least-squares fits. The 68\,per
cent confidence intervals for the parameters of these relationships
and the binned measurements were all calculated using a simple
bootstrap method: 10$^4$ bootstrap samples were formed by drawing,
with replacement, random values from the observed distribution, and
the distribution of the mean in the bootstrap samples provided the
confidence intervals. \Tref{tab:res} also presents results from
Spearman rank (i.e.~non-parametric) correlation tests between the
parameters of interest, including the correlation coefficient, $r$,
and probability, $p$, of the observed correlation being due to chance
under the null hypothesis of no correlation.

\begin{table*}
\begin{center}
  \caption{Main numerical results for correlation tests and linear
    fits between the individual \EBV\ values, measured in
    milli-magnitudes (mmag), and absorption redshift (\zab), neutral
    hydrogen column density (\NHI), the \Ion{Si}{ii} $\lambda$1526
    rest-frame equivalent width [\EWSi], the corresponding
    \Ion{Si}{ii} metallicity ($Z_{\rm Si}$) and column density
    (\Nion{Si}{ii}) derived using known scaling relations (for DLAs
    only; see \Sref{ssec:EBV_metals}), and the \Ion{C}{ii} equivalent
    width [\EWC]. The correlations and fits are performed for both SMC
    and LMC-like dust models using both the DLA sample and absorber
    sample [which includes sub-DLAs with $\lNHI\ge20.0$]. The first
    column describes the `model' being tested or fitted in each
    row. The Spearman rank correlation coefficient, $r$, and
    associated probability, $p$, are provided for the correlation
    tests.}
\label{tab:res}\vspace{-0.5em}
\begin{tabular}{llllll}\hline
\EBV\,[mmag]                       & \multicolumn{2}{c}{DLAs: $\lNHI\ge20.3$}                          && \multicolumn{2}{c}{All absorbers: $\lNHI\ge20.0$}                 \\
\multicolumn{1}{c}{Model}          & \multicolumn{1}{c}{SMC dust}    & \multicolumn{1}{c}{LMC dust}    && \multicolumn{1}{c}{SMC dust}    & \multicolumn{1}{c}{LMC dust}    \\\hline
$\left<\EBV\right>$                & \multicolumn{1}{c}{$3.0\pm1.0$} & \multicolumn{1}{c}{$6.2\pm2.2$} && \multicolumn{1}{c}{$3.6\pm0.9$} & \multicolumn{1}{c}{$7.2\pm1.9$} \medskip\\
\zab~corr.~test                    & $r=-0.03$, $p=0.35$             & $r=-0.03$, $p=0.40$             && $r=-0.00$, $p=0.99$             & $r=-0.01$, $p=0.87$             \\
$a+b\,\zab$                        & $a=8.0 \pm 7.8$,                & $a=11 \pm 19$,                  && $a=3.5 \pm 7.0$,                & $a=5 \pm 17$,                   \\
                                   & $b=-1.7 \pm 2.8$                & $b=-1.9 \pm 7.0$                && $b=0.0 \pm 2.4$                 & $b=0.7 \pm 6.2$                 \medskip\\
\NHI~corr.~test                    & $r= 0.04$, $p=0.24$             & $r= 0.05$, $p=0.21$             && $r= 0.02$, $p=0.51$             & $r= 0.02$, $p=0.43$             \\
$b\,[\NHI/{\rm cm}^{-2}]$          & $b=(3.5 \pm 1.0)\times10^{-21}$ & $b=(7.8 \pm 2.1)\times10^{-21}$ && $b=(3.8 \pm 1.0)\times10^{-21}$ & $b=(8.3 \pm 2.1)\times10^{-21}$ \\
$a+b\,[\NHI/{\rm cm}^{-2}]$        & $a=1.7 \pm 1.4$,                & $a=2.8 \pm 3.3$,                && $a=2.6 \pm 1.1$,                & $a=4.6 \pm 2.4$,                \\
                                   & $b=(2.4 \pm 1.4)\times10^{-21}$ & $b=(6.0 \pm 3.1)\times10^{-21}$ && $b=(1.9 \pm 1.3)\times10^{-21}$ & $b=(4.9 \pm 2.9)\times10^{-21}$ \medskip\\
\EWSi~corr.~test                   & $r=0.19$, $p=3.9\times10^{-4}$  & $r=0.19$, $p=3.8\times10^{-4}$  && $r=0.17$, $p=3.7\times10^{-4}$  & $r= 0.17$, $p=5.3\times10^{-4}$ \\
$b\,[\EWSi/{\rm \AA}]$             & $b=5.0 \pm 1.4$                 & $b=9.2 \pm 3.0$                 && $b=6.4 \pm 1.4$                 & $b=12 \pm 3$                    \\
$a+b\,[\EWSi/{\rm \AA}]$           & $a=-3.7 \pm 2.2$,               & $a=-11 \pm 5$,                  && $a=-1.8 \pm 2.2$,              & $a=-3.8 \pm 4.3$,               \\
                                   & $b=7.8 \pm 2.2$                 & $b=17 \pm 5$                    && $b=7.9 \pm 2.3$                 & $b=16 \pm 5$                    \\
$b\,Z_{\rm Si}$                    & $b=33  \pm 9$                   & $b=65 \pm 21$                   &&                                 &                                 \medskip\\
$\Nion{Si}{ii}$~corr.~test         & $r=0.15$, $p=4.2\times10^{-3}$  & $r=0.16$, $p=2.7\times10^{-3}$  &&                                 &                                 \\
$b\,[\Nion{Si}{ii}/{\rm cm}^{-2}]$ & $b=(5.2 \pm 1.3)\times10^{-16}$ & $b=(10 \pm 3)\times10^{-16}$    &&                                 &                                 \medskip\\
\EWC~corr.~test                    & $r=0.16$, $p=4.6\times10^{-3}$  & $r=0.16$, $p=5.7\times10^{-3}$  && $r=0.12$, $p=0.02$              & $r=0.11$, $p=0.03$              \\
$b\,[\EWC/{\rm \AA}]$              & $b=3.1 \pm 1.3$                 & $b=6.4 \pm 2.6$                 && $b=3.9 \pm 1.5$                 & $b=6.7 \pm 2.7$                 \\
$a+b\,[\EWC/{\rm \AA}]$            & $a=-4.2 \pm 2.3$,               & $a=-9.4 \pm 4.8$,               && $a=-2.8 \pm 2.4$,               & $a=-6.5 \pm 5.0$,               \\
                                   & $b=6.0 \pm 2.3$                 & $b=13 \pm 4$                    && $b=5.9 \pm 2.5$                 & $b=11 \pm 5$                    \\\hline
\end{tabular}
\end{center}
\end{table*}

\subsection{Shift in spectral index}\label{ssec:beta}

\begin{figure}
\begin{center}
\includegraphics[width=1.0\columnwidth]{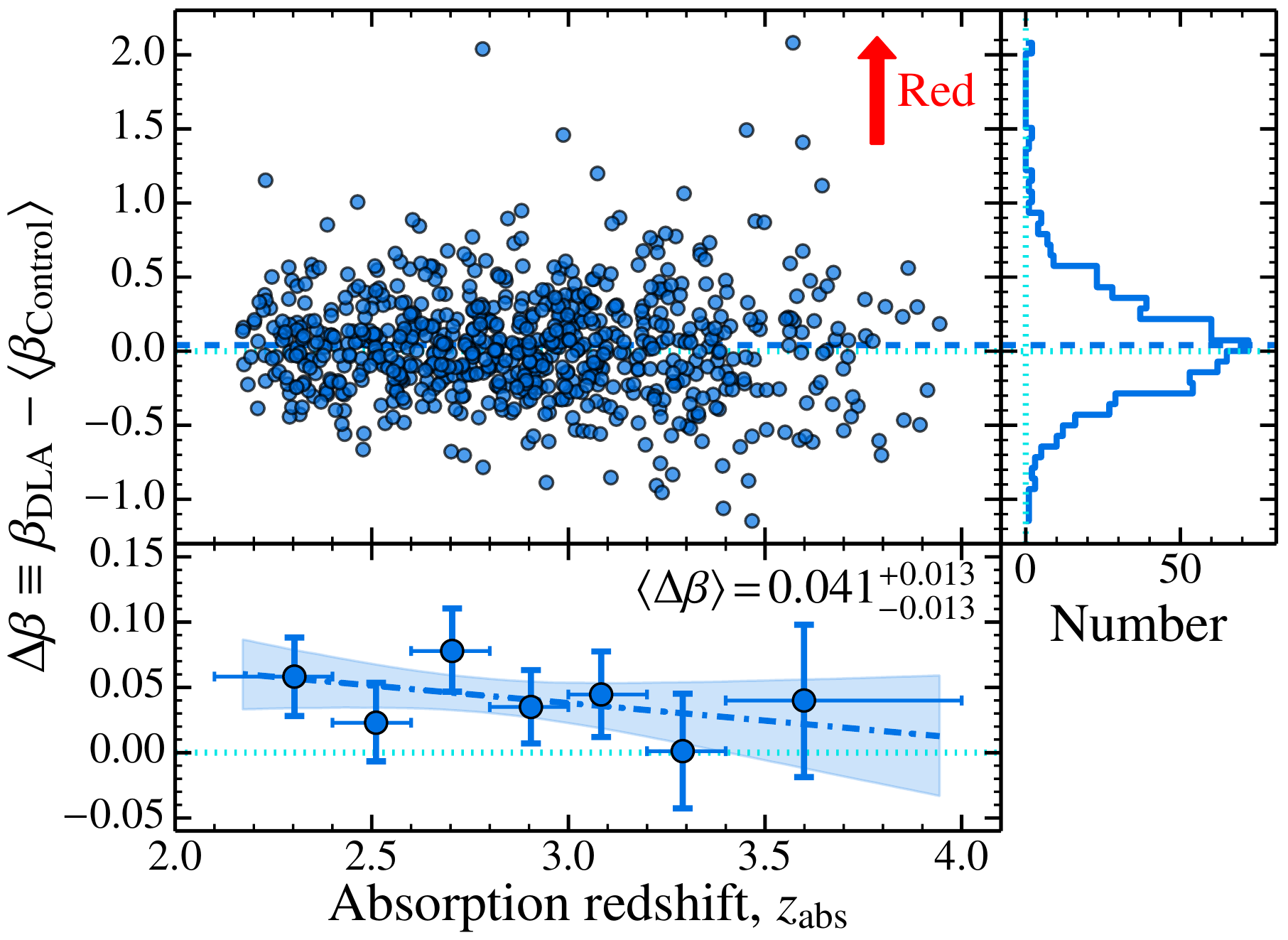}\vspace{-0.5em}
\caption{Distribution of the spectral index difference between the DLA
  quasars and their corresponding control quasar samples,
  $\Delta\beta$ [see \Eref{eq:Deltabeta}]. The upper panels show the
  overall $\Delta\beta$ distribution (histogram in right-hand panel)
  and its relationship with absorption redshift (left-hand upper
  panel). The dashed line shows the mean $\Delta\beta$ value. The
  lower panel shows the mean $\Delta\beta$ in absorption redshift bins
  (defined in \Fref{fig:redshift_dist}) with 68\,per cent bootstrap
  confidence intervals. The horizontal error bars indicate a bin's
  redshift range and the point is plotted at the mean absorber
  redshift within the bin. The mean $\Delta\beta$, averaged over all
  redshifts (with 68\,per cent confidence intervals), is also
  provided. The dot-dashed line is the best-fitting straight line to
  the individual $\Delta\beta$ values and the shaded region shows the
  68\,per cent bootstrap confidence region of the fit. Despite the
  weak redshift evolution in $\beta$ evident in \Fref{fig:beta_dist},
  the differential analysis used to derive $\Delta\beta$ here reveals
  no residual evolution in $\Delta\beta$.}
\label{fig:Deltabeta_dist}
\end{center}
\end{figure}

The upper panel of \Fref{fig:Deltabeta_dist} shows the measurements of
$\Delta\beta$ [\Eref{eq:Deltabeta}] for the DLA quasars against the
DLA absorption redshift. Compared to the $\beta$ values shown in
\Fref{fig:beta_dist}, it is immediately clear that there is no
apparent redshift evolution in $\Delta\beta$. However, the
distribution of $\Delta\beta$ values, shown in the histogram, has a
similar non-Gaussian, asymmetric shape as the $\beta$ values, as
expected.

The $\Delta\beta$ distribution in \Fref{fig:Deltabeta_dist} also shows
that any DLA dust reddening causes a much smaller shift in $\beta$
than the typical width of the distribution. Indeed, the mean
$\Delta\beta$ is only $0.041\pm0.013$, representing some evidence for
very low DLA dust reddening. The bootstrap distribution indicates
that $\Delta\beta$ is $>$0 at 3.2-$\sigma$ significance; that is, the
null hypothesis of no DLA dust reddening is ruled out at 99.7\,per
cent confidence. The lower panel of \Fref{fig:Deltabeta_dist} shows
the mean $\Delta\beta$ within the absorption redshift bins defined in
\Fref{fig:redshift_dist}. The binned results do not appear to reveal
any evidence for evolution in $\Delta\beta$ with redshift. Indeed, a
linear least-squares fit to the individual $\Delta\beta$ values, with
bootstrap estimates of the 1-$\sigma$ parameter uncertainties, yields
a best fitted relationship of
$\Delta\beta=(0.12\pm0.11)+\zab\times(-0.03\pm0.04)$, confirming this
impression.

Conducting the same analysis for the absorber sample [i.e.~including
sub-DLAs with $\lNHI\ge20.0$] yields a mean shift in the spectral
index of $\Delta\beta=0.047\pm0.012$, with $\Delta\beta>0$ at
4.8-$\sigma$ significance. That is, including the sub-DLAs somewhat
increases the mean $\Delta\beta$ and substantially increases the
significance of the reddening detection. As with the DLA sample, we do
not find any evidence for evolution in $\Delta\beta$ with redshift.

\subsection{Mean reddening and redshift evolution}\label{ssec:EBV_zabs}

\begin{figure*}
\begin{center}
\vbox{
 \centerline{\hbox{
   \includegraphics[width=0.9\columnwidth]{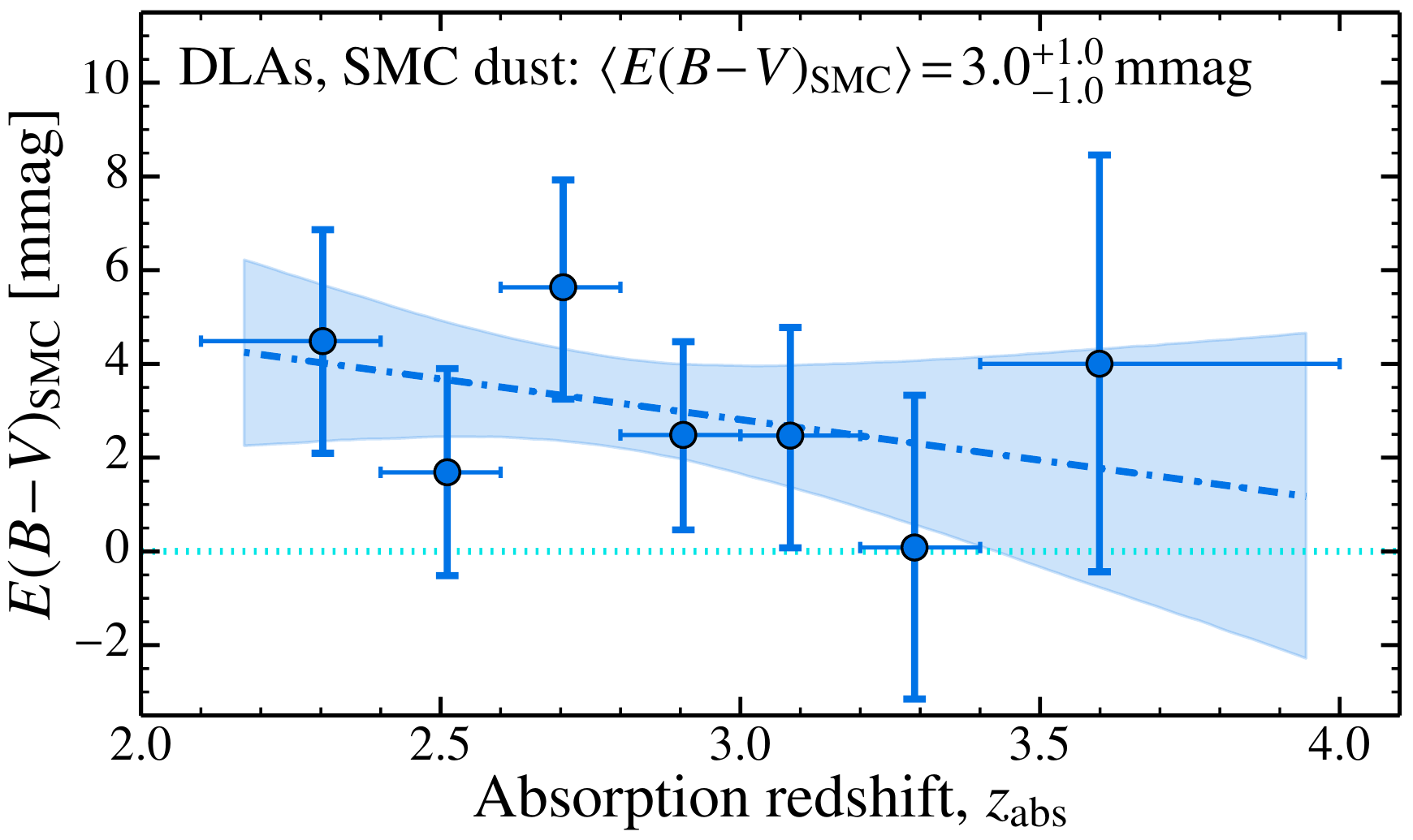}
   \includegraphics[width=0.9\columnwidth]{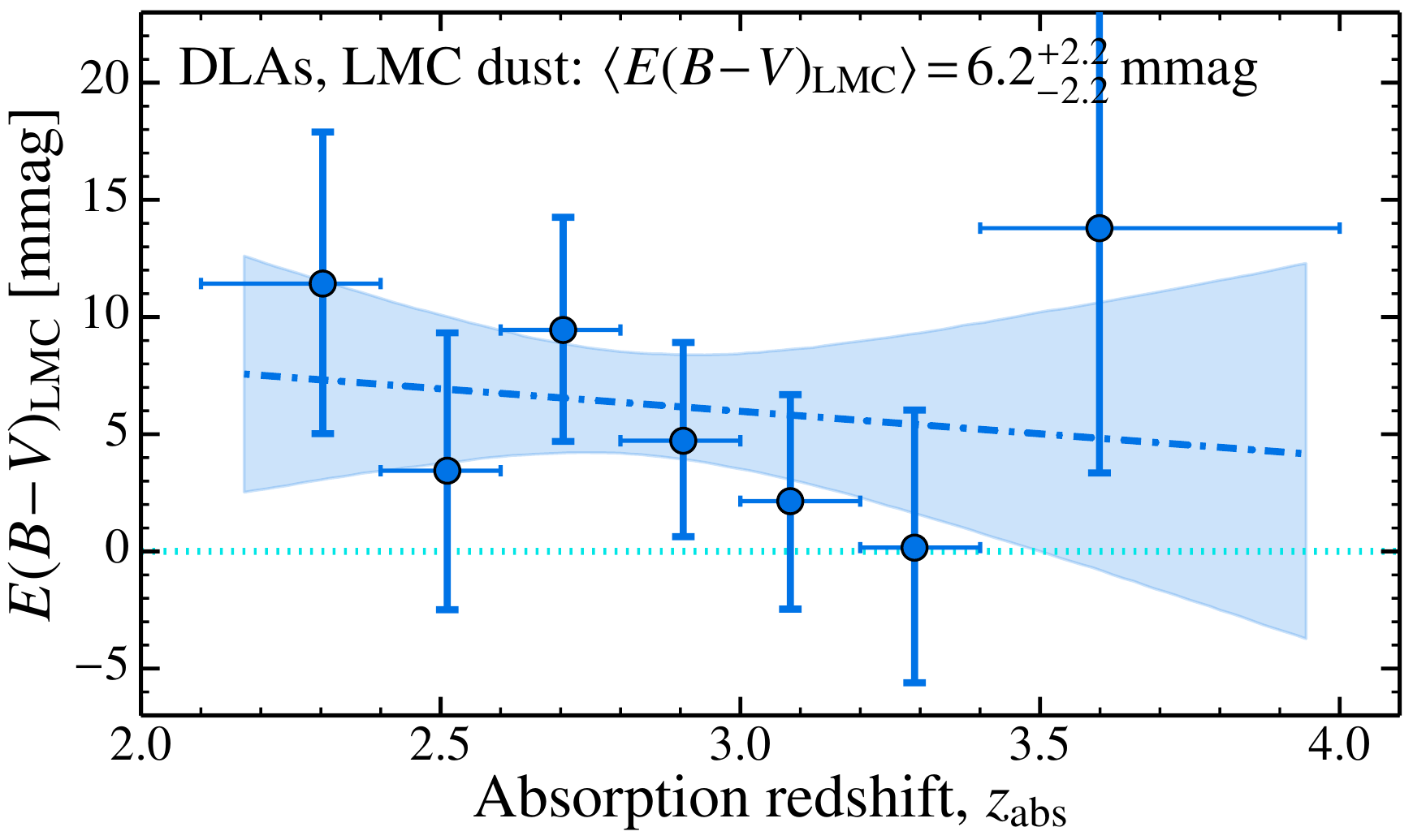}
 }}
 \centerline{\hbox{
   \includegraphics[width=0.9\columnwidth]{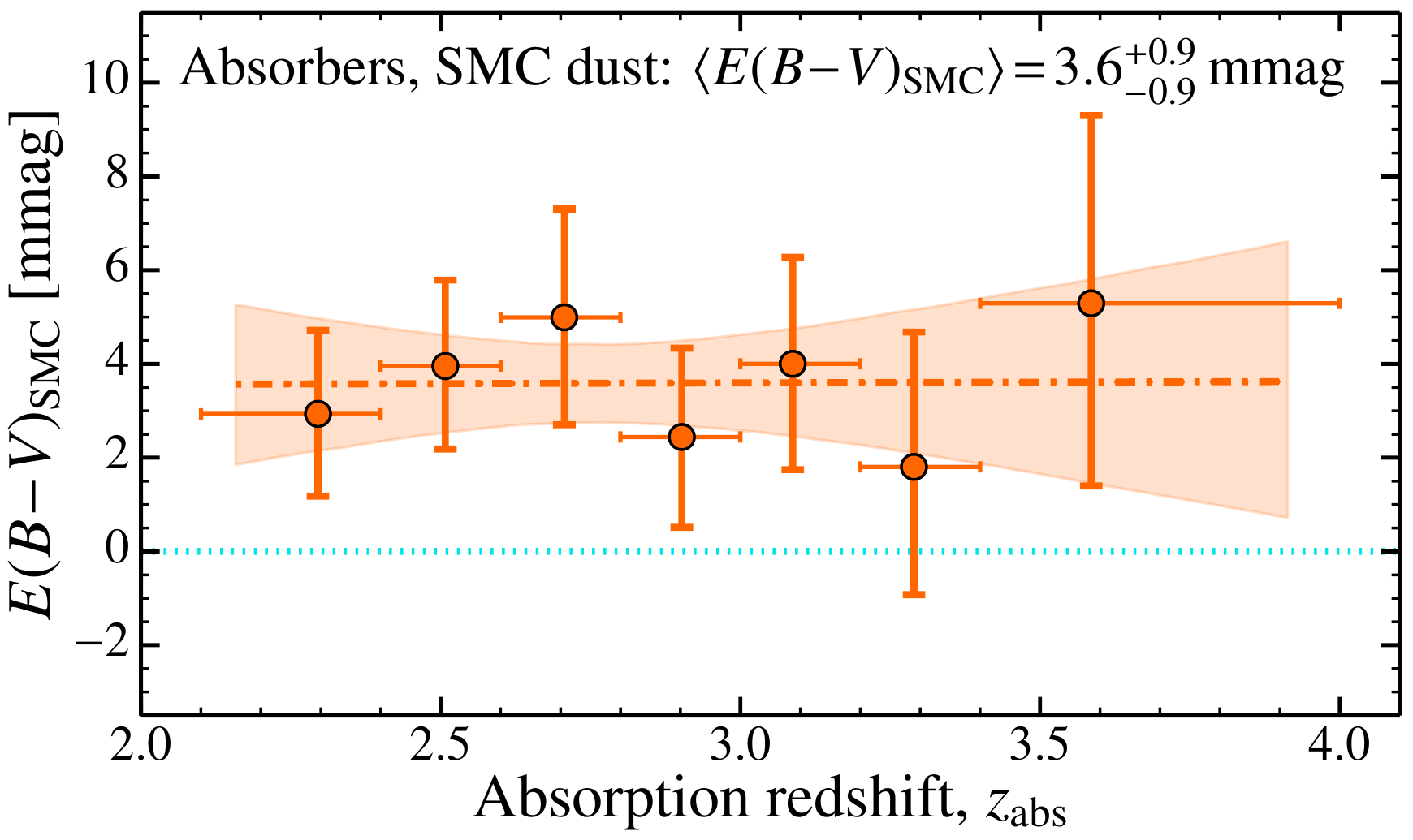}
   \includegraphics[width=0.9\columnwidth]{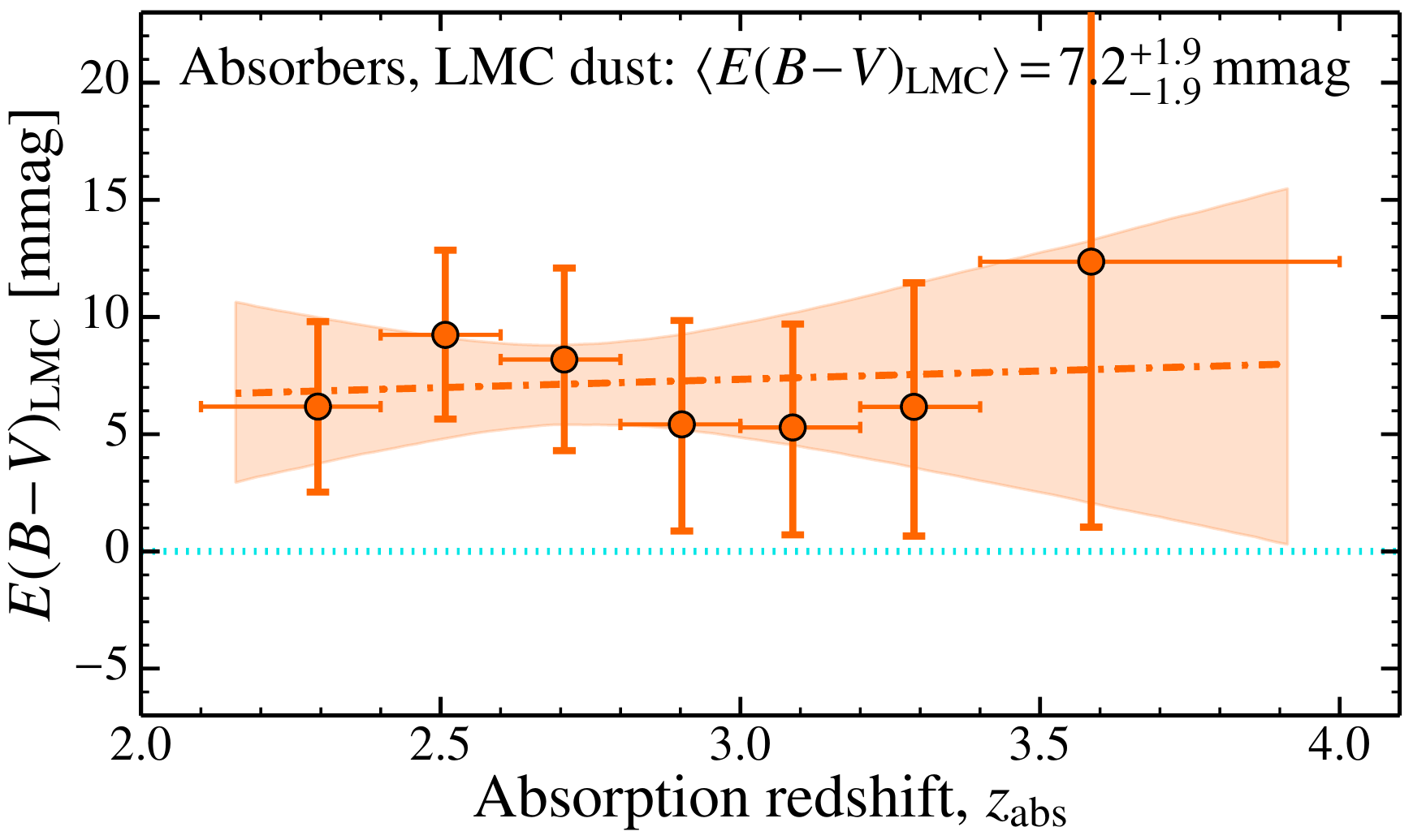}
 }}
}\vspace{-0.5em}
\caption{The mean colour excess, \EBV, in the absorber rest-frame in
  bins of absorption redshift (defined in
  \Fref{fig:redshift_dist}). The upper plots show the mean \EBVSMC\
  (left-hand plot) and \EBVLMC\ (right-hand plot) values for the DLAs,
  while the lower plots show the values for all absorbers with
  $\lNHI\ge20.0$. Note the doubled \EBV\ scale for the LMC plots. The
  vertical error bars represent the 68\,per cent bootstrap confidence
  intervals. The horizontal error bars indicate a bin's range and the
  point is plotted at the mean absorber redshift within the bin. The
  dot-dashed lines are the best-fit straight lines to the individual
  \EBV\ values as a function of redshift, with the 68\,per cent
  bootstrap confidence intervals in the fits shown as shaded regions.}
\label{fig:EBV_zabs}
\end{center}
\end{figure*}

\Fref{fig:EBV_zabs} shows the mean colour excess measurements, in the
redshift bins defined in \Fref{fig:redshift_dist}, after converting
each DLA (and absorber) quasar's $\Delta\beta$ value to \EBVSMC\ and
\EBVLMC, as described in \Sref{sec:analysis}. The Figures and
\Tref{tab:res} provide the mean \EBV, averaged over all redshifts, for
the DLA and absorber quasar samples in both dust models. For example,
assuming DLAs contain SMC-like dust, they cause a mean colour excess
of $\left<\EBVSMC\right>=3.0\pm1.0$\,mmag over the redshift range
$2<\zab<4$. Assuming the dust obeys an LMC-like dust law (i.e.~with a
shallower extinction curve), the corresponding mean colour excess is
higher, $\left<\EBVLMC\right>=6.2\pm2.2$\,mmag. The overall evidence
for $\EBV>0$ is, of course, very similar to that for $\Delta\beta>0$:
3.2-$\sigma$ significance, averaged over all redshifts, for both SMC
and LMC-like dust.

\Tref{tab:res} shows that both the colour excess and statistical
significance of \EBV\ increase when sub-DLAs with $\lNHI\ge20.0$ are
included (as also seen for $\Delta\beta$). In both dust models,
$\EBV=0$ is ruled out at 4.8-$\sigma$ significance.

\Fref{fig:EBV_zabs} reveals no evidence for redshift evolution in the
colour excess caused by dust in DLAs and sub-DLAs. The bootstrap
uncertainty in the (assumed) linear relationship between \EBV\ and
\zab\ is illustrated by the shaded regions in the Figures; these are
all clearly consistent with no evolution in \EBV. The best-fit
parameter values and uncertainties are provided in \Tref{tab:res}. For
example, the 1-$\sigma$ limit on evolution in \EBVSMC\ is 2.8\,mmag
per unit redshift for DLAs, assuming an SMC-like dust model, and a
similar value, 2.4\,mmag per redshift interval, when including
sub-DLAs. The lack of redshift evolution is also confirmed in
\Tref{tab:res} with the Spearman rank tests for correlations. These
return $p$ values $>$35\,per cent in all four cases (DLA and absorber
samples with SMC and LMC-like dust), indicating no significant
correlation between \EBV\ and \zab. Over the same redshift interval,
$\zab=4$--2, the mean DLA metallicity has been found to increase by a
factor of $\approx$2.5 \citep{Rafelski:2012:89,Jorgenson:2013:482}. If
the mean dust column scales linearly with DLA metallicity, as expected
in simple models \citep[e.g.][]{Vladilo:2005:461}, we should expect a
slope of $b\approx1.3$\,mmag per unit redshift in \EBVSMC. However,
\Tref{tab:res} shows that our 1-$\sigma$ sensitivity to a slope is
$b=2.8$\,mmag per unit redshift. That is, the SDSS DLA dataset is not
large enough to detect the evolution in dust content expected from the
observed metallicity evolution in DLAs.

\subsection{Reddening vs.~neutral hydrogen column density}\label{ssec:EBV_NHI}

\begin{figure}
\begin{center}
\vbox{
 \includegraphics[width=1.0\columnwidth]{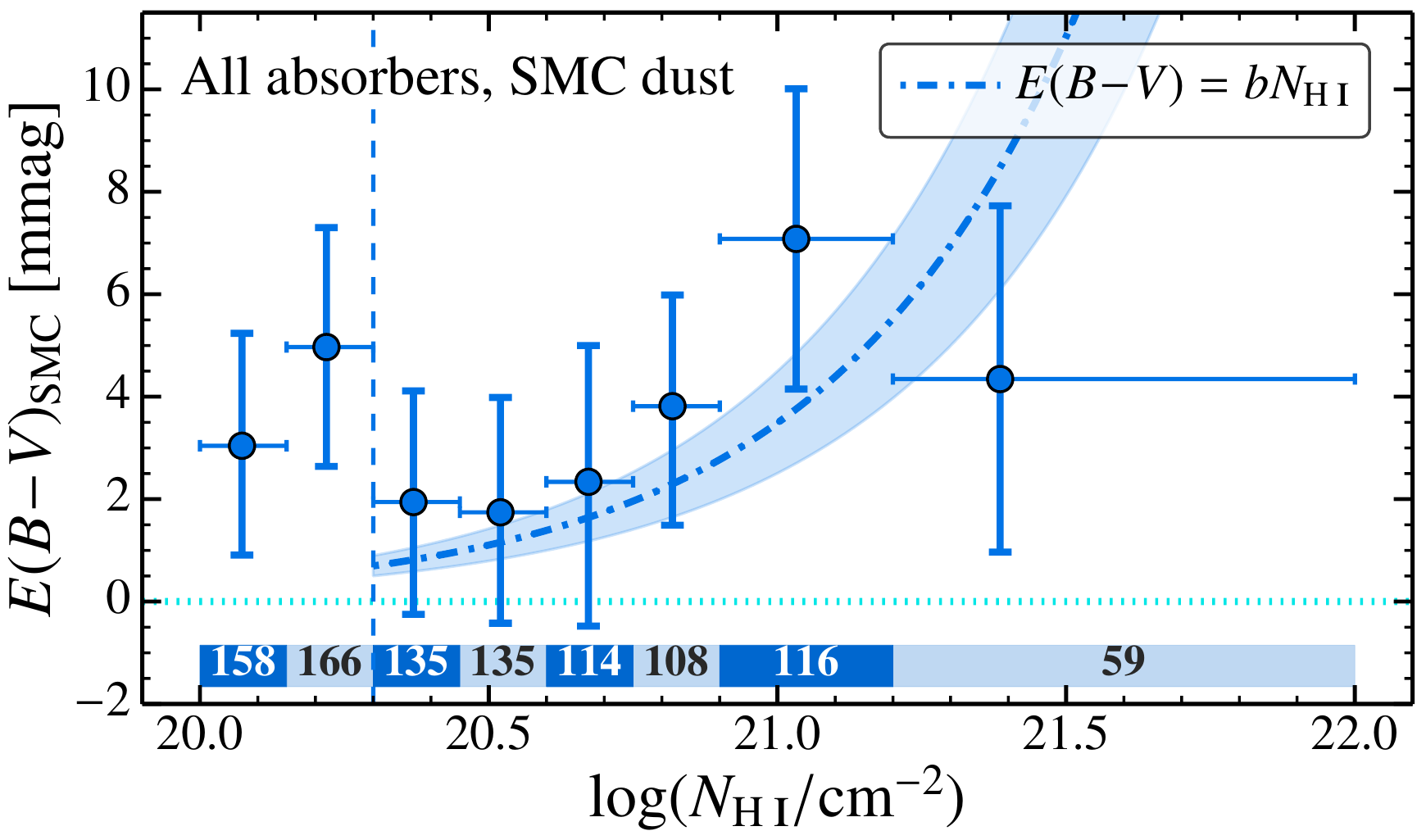}
 \includegraphics[width=1.0\columnwidth]{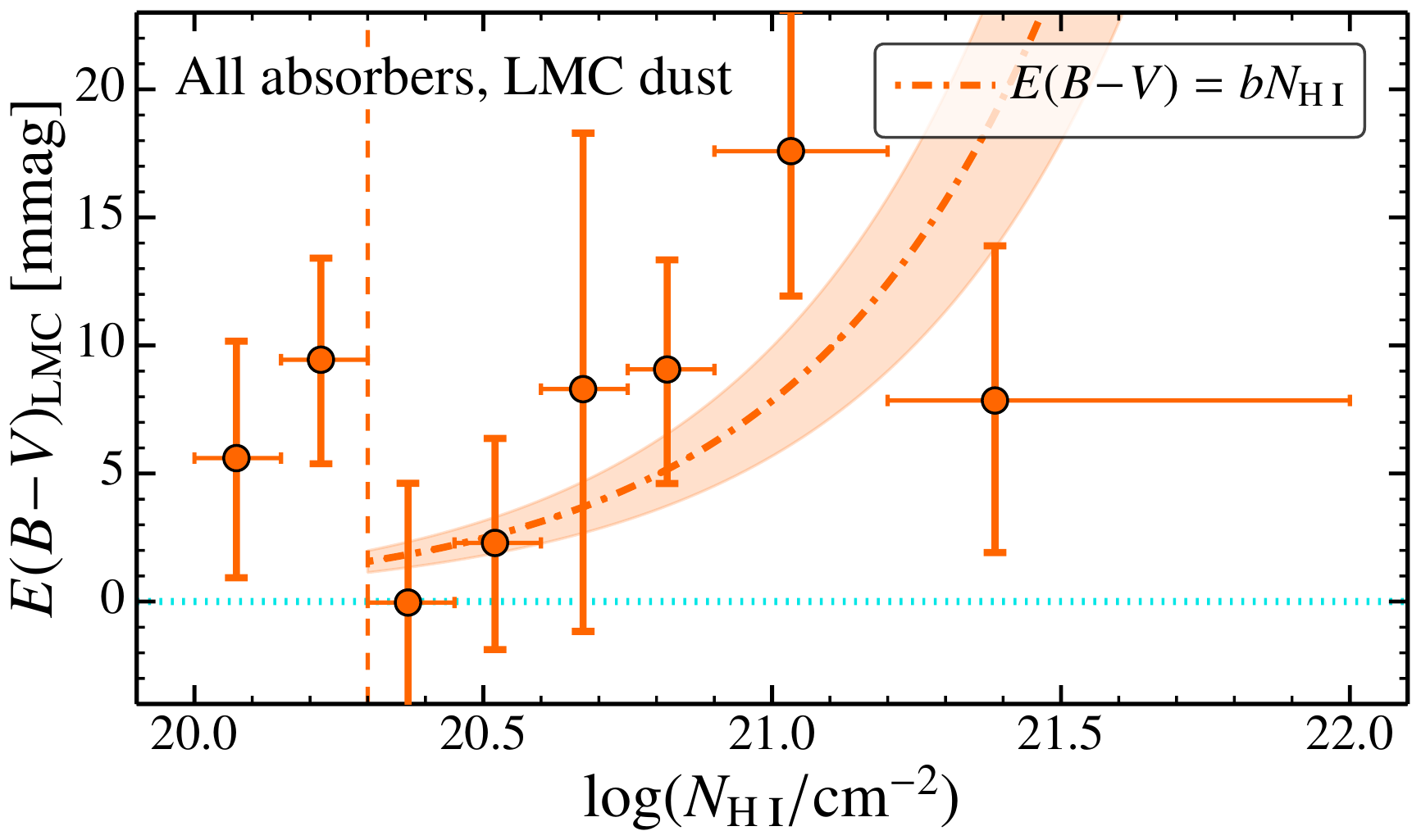}
}\vspace{-0.5em}
\caption{The mean colour excess, \EBV, in the absorber rest-frame in
  bins of $\log\NHI$. The upper plot shows the mean \EBVSMC\ values
  while the lower plot shows the \EBVLMC\ values. Note the doubled
  \EBV\ scale for the LMC plot. The shaded bar in the upper plot shows
  the number of absorbers in each $\log\NHI$ bin. The vertical error
  bars represent the 68\,per cent bootstrap confidence intervals. The
  horizontal error bars indicate a bin's range and the point is
  plotted at the mean $\log\NHI$ within the bin. The dot-dashed lines
  are the best-fit straight lines to the individual \EBV\ values as a
  function of \NHI\ [not $\log\NHI$, hence their curvature in the plot],
  with the 68\,per cent bootstrap confidence intervals in the fits
  shown as shaded regions. The dashed lines delineate DLAs and
  sub-DLAs. Note that the results in both plots were derived from
  analysing the absorber (and corresponding non-absorber) sample
  containing DLAs and sub-DLAs with $\lNHI\ge20.0$. The results from
  the DLA/non-DLA analysis are very similar for $\lNHI\ge20.3$.}
\label{fig:EBV_NHI}
\end{center}
\end{figure}

\Fref{fig:EBV_NHI} shows the mean colour excess measurements in bins
of $\log\NHI$\ using the sample of all absorbers with
$\lNHI\ge20.0$. Using the DLA sample returns very similar results at
$\lNHI\ge20.3$ but, by definition, does not extend to lower \NHI, so
we focus on the full absorber sample here. The binned results in
\Fref{fig:EBV_NHI} provide no evidence for strong variations in \EBV\
with increasing \NHI. Nor do the Spearman rank correlation tests in
\Tref{tab:res} indicate significant correlation or anti-correlation
($p>0.2$ in all cases). It is instead notable that the amount of
reddening is similar at all \Ion{H}{i} column densities over a 1.5
order-of-magnitude range and that the highest and lowest \NHI\ bins
show no discernable differences.

In both the SMC and LMC, the colour excess is correlated with \NHI,
with the reddening per H atom (including contributions from H$_2$)
being $\EBV/N_{\rm H}\approx2\times10^{-23}$ \citep{Martin:1989:219}
and ${\approx}4.5\times10^{-23}$\,mag\,cm$^2$
\citep{Fitzpatrick:1985:219}, respectively. This motivates a linear
fit between our individual \EBV\ results and \NHI, as shown by the
dot-dashed lines and 1-$\sigma$ shaded regions in
\Fref{fig:EBV_NHI}. Although there is no evidence for such a
relationship, as discussed above, the data in \Fref{fig:EBV_NHI} are
not inconsistent with the best-fit relationships derived. For better
comparison with the SMC and LMC, we restricted these fits to DLAs
according to the traditional threshold for selecting only the most
neutral absorbers [i.e.~$\lNHI\ge20.3$]. The resulting parameter
estimates are given in \Tref{tab:res}. Assuming a negligible H$_2$
molecular fraction in our DLAs, the mean reddening per H atom is
$\EBVSMC/N_{\rm H}\approx0.4\times10^{-23}$ and $\EBVLMC/N_{\rm
  H}\approx0.8\times10^{-23}$\,cm$^2$. These are only $\approx$20\,per
cent of the SMC and LMC values given above.

Even if the assumption of a simple linear relationship between \EBV\
and \NHI\ is valid for DLAs, the much smaller mean reddening per H
atom for DLAs is most likely explained by their much lower average
metallicity of $\sim$1/30 solar
\citep[e.g.][]{Rafelski:2012:89,Jorgenson:2013:482} compared to the
SMC and LMC \citep[$\sim$1/7 and $\sim$1/3
solar;][]{Kurt:1998:202,Draine:2003:241}. However, while the
interstellar medium within each of the Magellanic Clouds shows little
metallicity variation, the DLA population exhibits a very large range,
$\MH{M}\sim-3.0$--0.0\footnote{We use the standard notation for
  metallicity, with $\MH{M}\equiv\log[N({\rm M})/N({\rm
    H})]-\log[N({\rm M})/N({\rm H})]_\odot$ and $Z_{\rm
    M}\equiv10^{\left[{\rm M}/{\rm H}\right]}$, where the solar
  abundance ratios, $[N({\rm M})/N({\rm H})]_\odot$, are taken from
  \citet{Asplund:2009:481}.}. Therefore, even if \EBV\ correlates
tightly with \NHI\ in the different regions of individual high-redshift
galaxies (e.g.~disk, gaseous halo etc.) with $\lNHI\ge20.3$, the large
metallicity range of the DLAs -- an ensemble of different regions from
different galaxies -- will tend to flatten any \EBV\ vs.~\NHI\
correlation, possibly leading to an apparently flat relationship like
those observed in \Fref{fig:EBV_NHI}.

\subsection{Reddening vs.~metal line equivalent width, metallicity and metal column density}\label{ssec:EBV_metals}

In addition to measuring \NHI\ for each DLA candidate,
\citet{Noterdaeme:2009:1087} determined the equivalent widths of
several strong metal absorption lines when they fell redwards of the
quasar \lya\ emission line. \Ion{Si}{ii}\,$\lambda$1526 is the
strongest \Ion{Si}{ii} transition and its rest-frame equivalent width,
\EWSi, was reported for 355 of the 730 DLA quasars in our statistical
analysis (and 428 of the 991 absorber quasars). Similarly, \EWC\ was
reported for 303 DLA quasars (and 373 absorber quasars). The
unreported cases will be a mix of non-detections (i.e.~where the metal
absorption was too weak to detect) and where detection was precluded,
either because the transitions fell bluewards of the quasar \lya\
emission line or due to some artefact in the spectra (e.g.~bad
pixels).

\begin{figure}
\begin{center}
\vbox{
 \includegraphics[width=1.0\columnwidth]{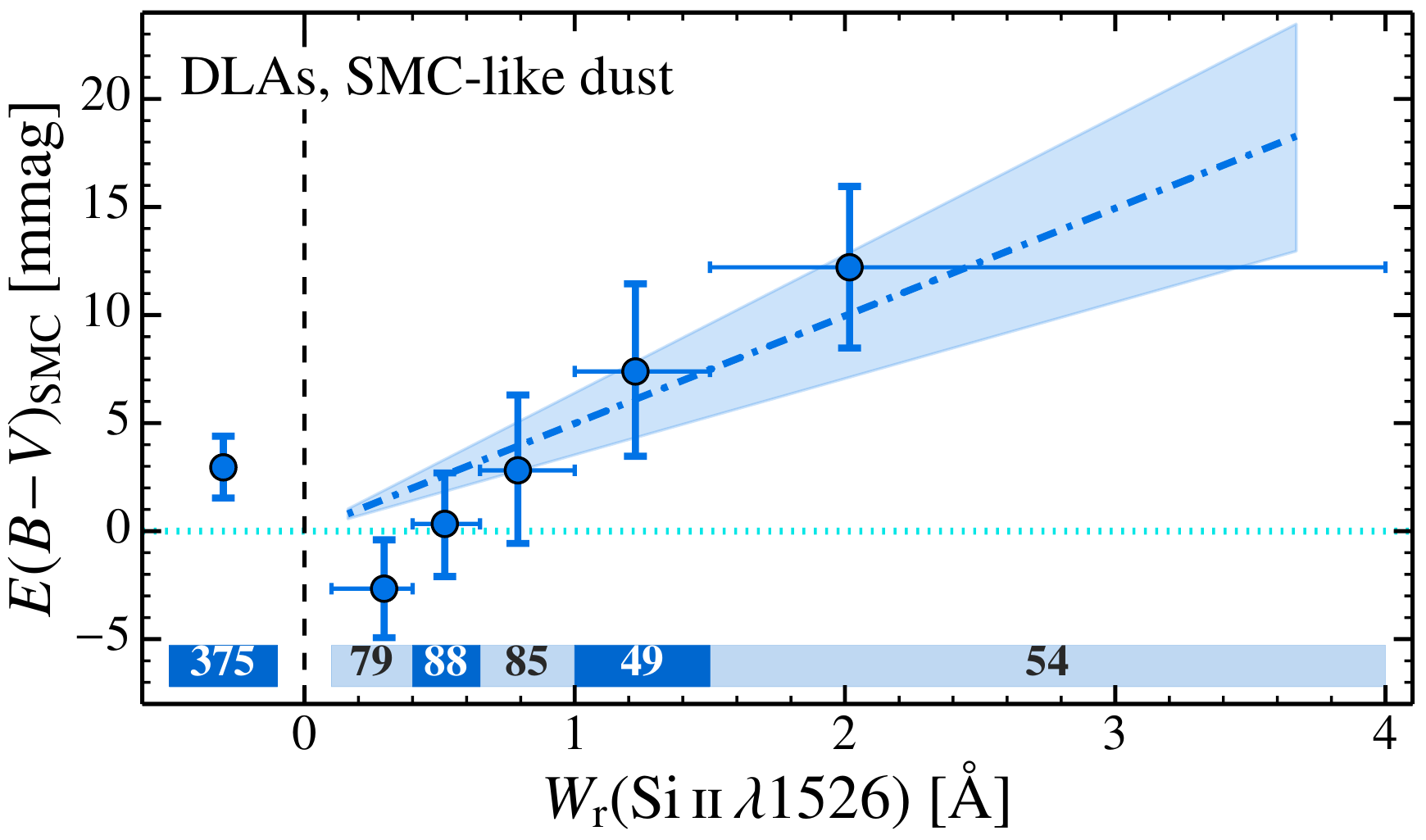}
 \includegraphics[width=1.0\columnwidth]{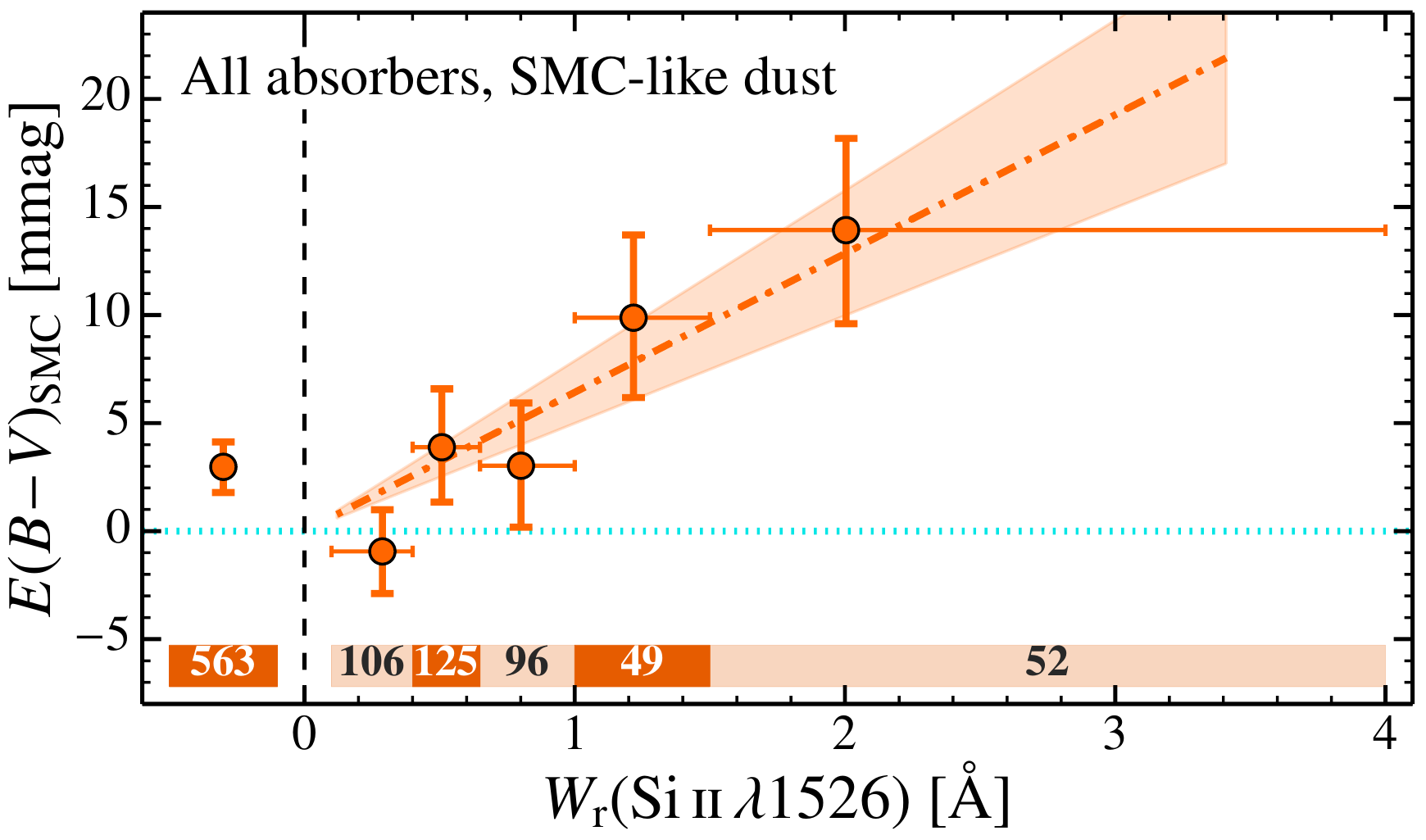}
}\vspace{-0.5em}
\caption{The mean colour excess for SMC-like dust, \EBVSMC, in the
  absorber rest-frame in bins of \Ion{Si}{ii}\,$\lambda$1526 rest
  equivalent width, \EWSi. The upper plot shows the mean \EBVSMC\
  values for the DLAs, while the lower plot shows the values for all
  absorbers with $\lNHI\ge20.0$. The shaded bar in each plot shows the
  number of absorbers in each bin. The left-most point in each plot is
  the mean \EBV\ value for DLAs/absorbers in which \Ion{Si}{ii} was
  not detected (see main text). The vertical error bars represent the
  68\,per cent bootstrap confidence intervals. The horizontal error
  bars indicate a bin's range and the point is plotted at the mean
  \EWSi\ within the bin. The dot-dashed lines are the best-fit
  straight lines to the individual \EBV\ values as a function of
  \EWSi, with the 68\,per cent bootstrap confidence intervals in the
  fits shown as shaded regions.}
\label{fig:EBV_EWSi}
\end{center}
\end{figure}

\begin{figure}
\begin{center}
\vbox{
 \includegraphics[width=1.0\columnwidth]{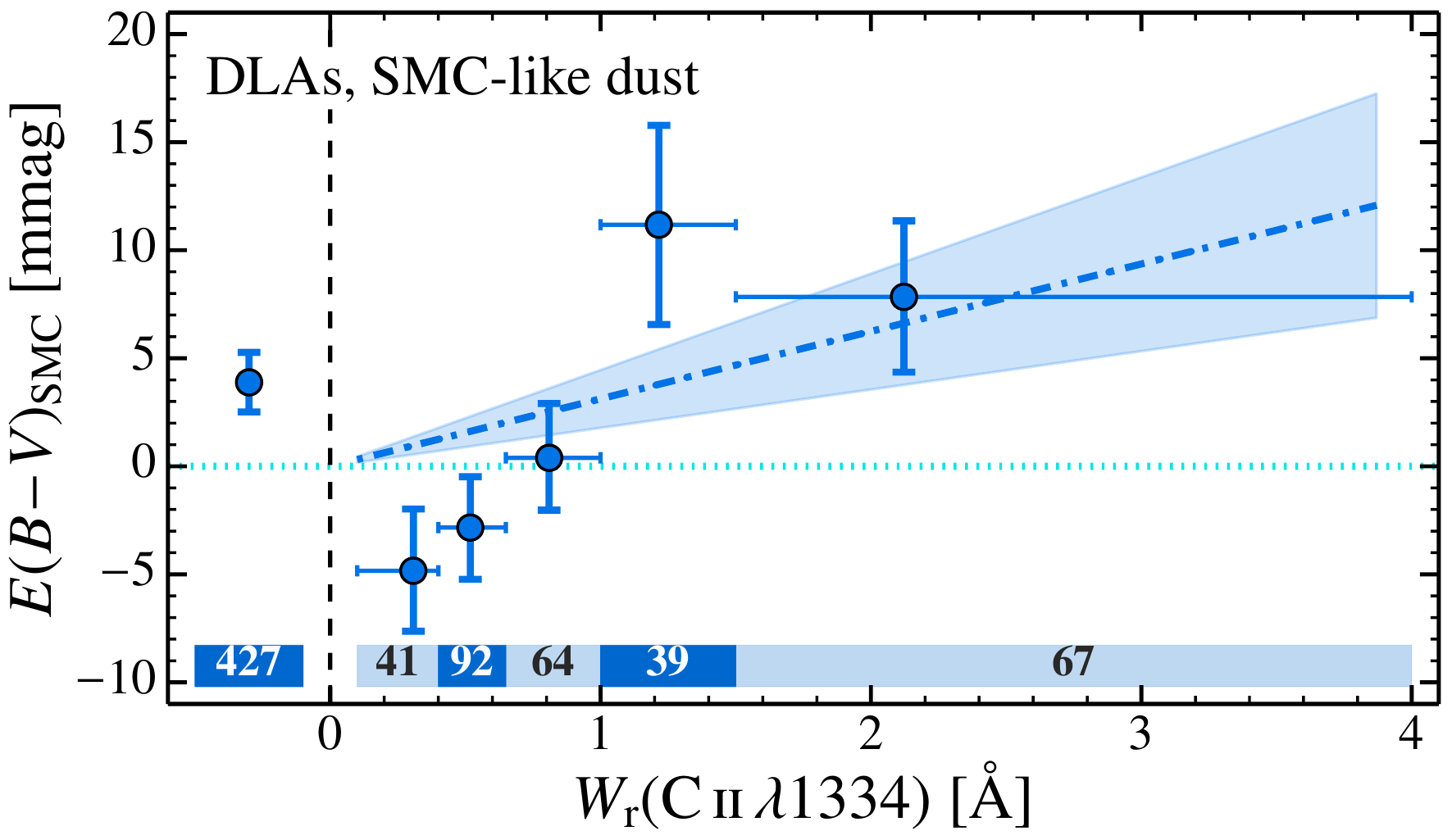}
 \includegraphics[width=1.0\columnwidth]{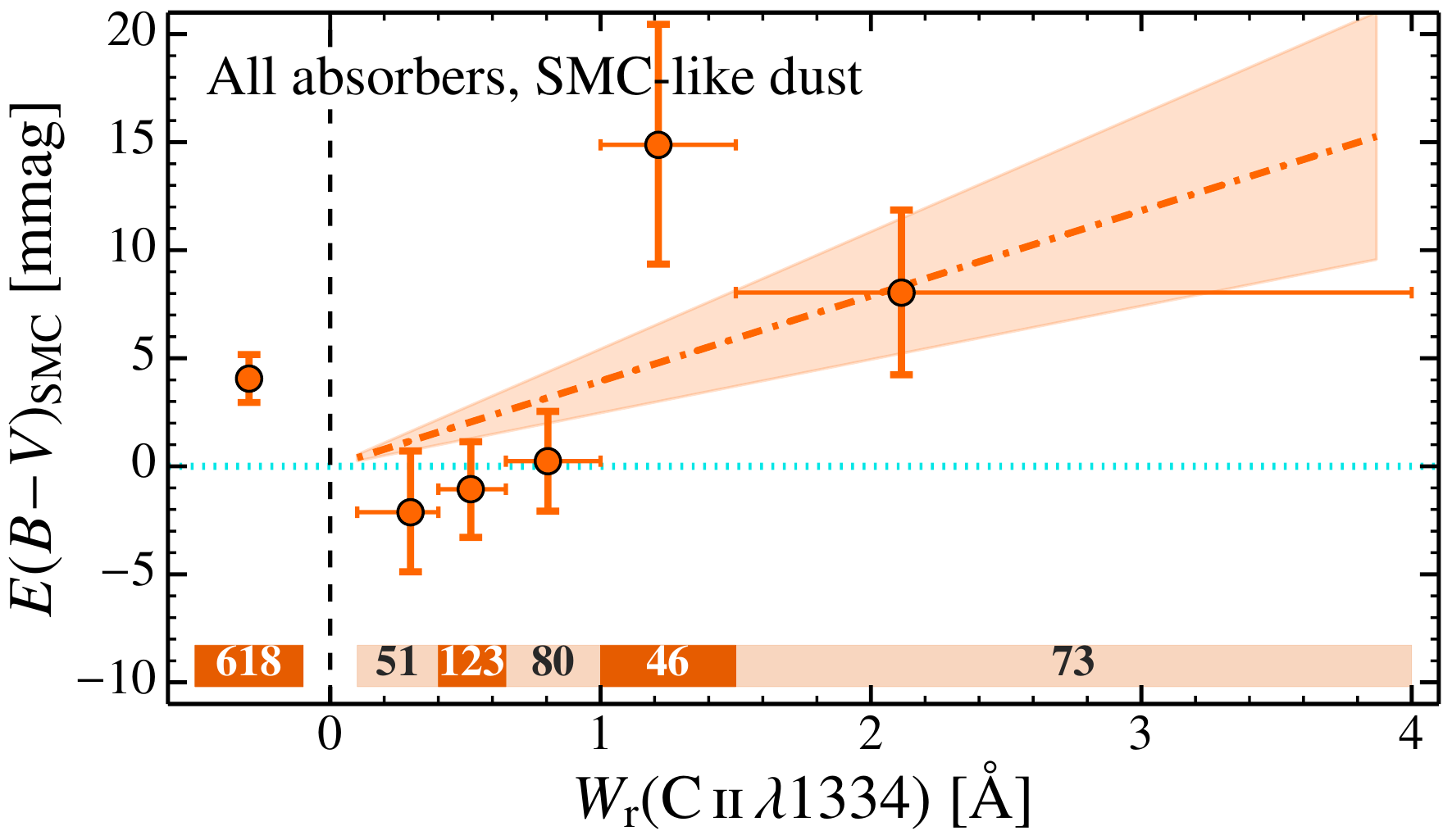}
}\vspace{-0.5em}
\caption{The mean colour excess for SMC-like dust, \EBVSMC, in the
  absorber rest-frame in bins of \Ion{C}{ii}\,$\lambda$1334 rest
  equivalent width, \EWC. The upper plot shows the mean \EBVSMC\ for
  DLAs while the lower plot shows the values for all absorbers with
  $\lNHI\ge20.0$. The shaded bars show the number of absorbers in
  each bin. The left-most point in each plot is the mean \EBVSMC\
  value for DLAs/absorbers in which \Ion{C}{ii} was not detected (see
  main text). The vertical error bars represent the 68\,per cent
  bootstrap confidence intervals. The horizontal error bars indicate a
  bin's range and the point is plotted at the mean \EWC\ within the
  bin. The dot-dashed lines are the best-fit straight lines to the
  individual \EBVSMC\ values as a function of \EWC, with the 68\,per
  cent bootstrap confidence intervals in the fits shown as shaded
  regions.}
\label{fig:EBV_EWC}
\end{center}
\end{figure}

Figures \ref{fig:EBV_EWSi} and \ref{fig:EBV_EWC} show the mean colour
excess measurements in bins of \EWSi\ and \EWC, respectively. There is
a clear tendency for systems with larger metal-line equivalent widths
to exhibit more reddening. The correlation tests in \Tref{tab:res}
confirm this: for example, \EBVSMC\ correlates strongly with \EWSi\
for DLAs, with the Spearman rank test giving a probability of
$p=0.04$\,per cent of the observed level of correlation occurring by
chance alone, i.e.~a $\approx$3.5-$\sigma$ significance. The
correlation is of similar strength, but lower significance
($\approx$2.8\,$\sigma$), between \EBVSMC\ and \EWC. Figures
\ref{fig:EBV_EWSi} and \ref{fig:EBV_EWC} also show the best linear
fits to the ensemble of \EBV\ and metal-line equivalent width
measurements from individual absorbers. \Tref{tab:res} provides the
best-fitting parameters: for example,
$\EBVSMC=(5.0\pm1.4)\times[\EWSi/$\AA$]$\,mmag for DLAs.

A strong correlation between DLA dust content and metal-line column
density is perhaps the most basic and assumption-free relationship we
should expect \citep[e.g.][]{Vladilo:2005:461}. Given the SDSS
spectral resolution of $\sim$170\,\kms\ and the typical rest-frame
equivalent width detection limits ($\sim$0.15\,\AA\ at 1-$\sigma$),
the detected metal transitions will be predominantly
saturated. Therefore, variation in their equivalent widths will more
directly reflect variations in the total velocity width of the
underlying absorption profile -- i.e.~the velocity spread amongst the
individual $\sim$5--10-\kms-wide components -- rather than the metal
column densities of those individual components. Nevertheless, a
broader profile will in general comprise more components, so the metal
column density should correlate strongly with equivalent width. That
is, we should expect the dust column density and \EBV\ to correlate
strongly with \EWSi\ and \EWC. Therefore, the increase in \EBV\ with
increasing metal-line equivalent width in Figs.~\ref{fig:EBV_EWSi} and
\ref{fig:EBV_EWC} indicates that the mean reddening of DLA quasars we
observe (\Fref{fig:Deltabeta_dist}) is really due to dust in the DLAs
and not due (primarily) to some spurious, systematic effect.

Considering all absorbers with $\lNHI\ge20.0$, the correlations for
\EBV\ and \EWSi\ have similar strengths and significance, and the
best-fit linear relationships have similar slopes, as for DLAs
(assuming either SMC or LMC-like dust). However, for \EWC, the
correlation becomes weaker and substantially less significant:
e.g.~for SMC-like dust, the by-chance correlation probability between
\EBVSMC\ and \EWC\ increases from $p=0.5$\,per cent for the 303 DLA
quasars to 2\,per cent for the 373 absorber quasars. Comparison of the
upper and lower panels of \Fref{fig:EBV_EWC}, which show the
\EBVSMC--\EWC\ relationships for DLA and absorber quasars, indicates
that the main difference is at low values of \EWC: for DLA quasars the
two lowest-\EWC\ bins lie $\approx$1.1 and 1.7-$\sigma$ below zero
reddening, while for the absorber quasars they are closer to
zero. That is, the strength of the \EBV--\EWC\ correlation for DLAs
may be overestimated in the DLA and absorber quasar samples. While the
deviation of \EBV\ below zero at low \EWC\ is consistent with a
statistical fluctuation, it may indicate that we generally
underestimate \EBV\ in our sample. We consider one possible
systematic effect that could, in principle, cause this -- the SDSS
colour-selection algorithm for selecting quasars for spectroscopic
follow-up -- in \Sref{sec:bias}, though we conclude there that it
causes $<$5\,per cent bias at all redshifts in our samples.

\begin{figure}
\begin{center}
\vbox{
 \includegraphics[width=1.0\columnwidth]{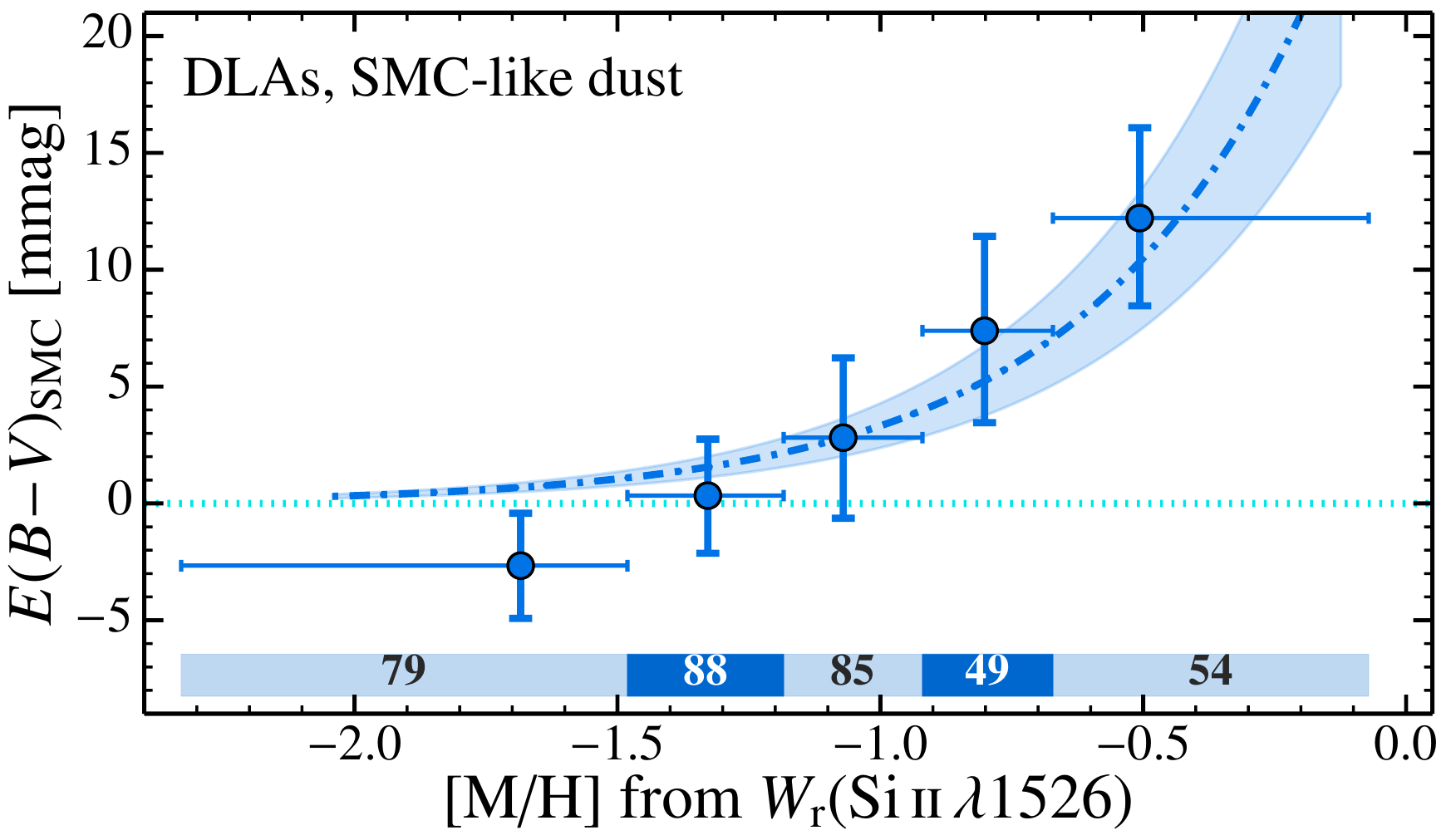}
 \includegraphics[width=1.0\columnwidth]{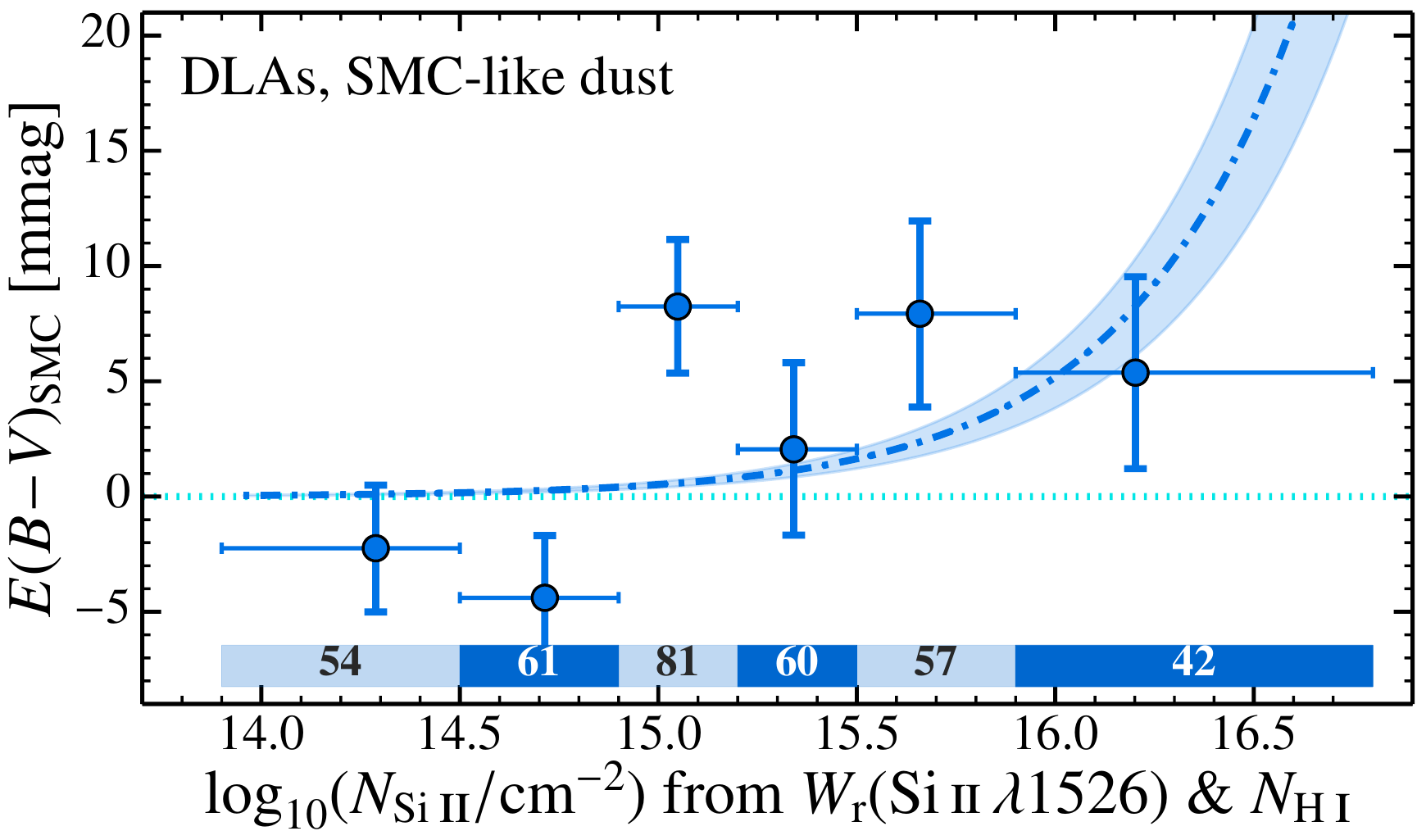}
}\vspace{-0.5em}
\caption{Relationship between mean colour-excess for SMC-like dust in
  DLAs and the metallicity, [M/H], and \Ion{Si}{ii} column density,
  \Nion{Si}{ii}, inferred from \EWSi\ and \NHI. The upper panel shows
  \EBVSMC\ in bins of [M/H] and the best-fitted linear relationship
  between \EBVSMC\ and \EWSi\ (dot-dashed line), both converted to
  metallicity using the known, tight correlation between \EWSi\ and
  [M/H] in DLAs in \Eref{eq:metallicity}. The shaded bars show the
  number of absorbers in each bin. DLAs for which \EWSi\ was not
  reported by \citet{Noterdaeme:2009:1087} are excluded. The vertical
  error bars represent the 68\,per cent bootstrap confidence
  intervals. The horizontal error bars indicate a bin's range and the
  point is plotted at the mean [M/H] within the bin. The lower panel
  shows the same information for bins of \Nion{Si}{ii} derived from
  \EWSi\ using \Eref{eq:metallicity} and the \NHI\ value for each
  DLA. The dot-dashed line shows the best-fit straight line to the
  individual \EBVSMC\ values as a function of \Nion{Si}{ii}, with the
  68\,per cent bootstrap confidence interval in the fit shown as
  a shaded region.}
\label{fig:EBV_MSi}
\end{center}
\end{figure}

The equivalent width of \Ion{Si}{ii}\,$\lambda$1526 is also
particularly interesting because it has been found to correlate very
strongly with DLA metallicity, [M/H], with only a $\sim$0.25\,dex
scatter and a linear relationship as follows \citep{Prochaska:2008:59}:
\begin{equation}\label{eq:metallicity}
[{\rm M}/{\rm H}]=(-0.92\pm0.05) + (1.41\pm0.10)\log_{10}[\EWSi/{\rm \AA}]\,.
\end{equation}
Similarly tight and strong relationships were also reported by
\citet{Kaplan:2010:619}, \citet{Neeleman:2013:54} and
\citet{Jorgenson:2013:482}. The upper panel of \Fref{fig:EBV_MSi}
shows the result of converting \EWSi\ to metallicity using
\Eref{eq:metallicity} for the DLAs and the linear fit between \EBVSMC\
and \EWSi\ from \Fref{fig:EBV_EWSi}'s upper panel. \Tref{tab:res} also
shows the results for the best linear fits between the ensemble of
\EBV\ values for DLAs and $Z_{\rm Si}\equiv10^{[{\rm M}/{\rm H}]}$
derived using \Eref{eq:metallicity}; for example,
$\EBVSMC=(33\pm9)\,Z_{\rm Si}$\,mmag. By comparison, assuming that
\EBV\ is instead linearly proportional to \EWSi, \Eref{eq:metallicity}
implies that \EBV\ should be proportional to $Z_{\rm Si}^{1/1.41}$;
for example, from the fitting results in \Tref{tab:res} for SMC-like
dust in DLAs, the relationship would be $\EBVSMC=(23\pm6)\,Z_{\rm
  Si}^{1/1.41}$\,mmag. While both relationships appear to reasonably
reflect the underlying trend between \EBV\ and metal-line equivalent
width and/or metallicity that we observe, the significant scatter in
our \EBV\ values clearly prevents us from assessing which relationship
is a more accurate description.

The lower panel of \Fref{fig:EBV_MSi} shows the result of converting
\EWSi\ to a \Ion{Si}{ii} column density using \Eref{eq:metallicity}
and the neutral hydrogen column density for each DLA,
i.e.~$\Nion{Si}{ii}=10^{-0.92}\left[\Nion{Si}{ii}/\NHI\right]_\odot\EWSi^{1.41}\NHI$. \Tref{tab:res}
shows that, while a correlation between \EBV\ and \Nion{Si}{ii} is
apparent, the statistical significance is lower than that between \EBV\
and \EWSi. For example, \EBVSMC\ is correlated with \EWSi\ at
$\approx$3.5-$\sigma$ but only at $\approx$2.8-$\sigma$ with the
values of \Nion{Si}{ii} derived in this way. This is likely due to the
lack of any strong correlation between \EBV\ and \NHI\ in our sample
(see \Fref{fig:EBV_NHI} and \Tref{tab:res}). The best-fitted linear
relationship between the individual \EBV\ measurements and the
\Nion{Si}{ii} values is provided in \Tref{tab:res} and, for SMC-like
dust in DLAs, plotted in \Fref{fig:EBV_MSi}. The reduced significance
of the correlation is apparent, with a larger scatter around the
best-fit relationship observed here compared to, e.g., the
relationship with [M/H] in the upper panel.

\section{Magnitude and colour selection biases}\label{sec:bias}

Dust in DLAs will cause both extinction and reddening of DLA
quasars. Therefore, it is important to address the possible effect
that the magnitude and colour-selection criteria of SDSS quasars
\citep{Richards:2002:2945} may have on the measured reddening
signal. Dusty DLAs may be preferentially selected or rejected, with
the degree of bias varying with redshift. This possibility has not
been thoroughly addressed in previous SDSS DLA reddening studies,
possibly because the selection algorithm is very complex and not
available in portable, public software. The colour-selection criteria
are highly redshift-dependent, especially around redshifts where
optical quasar colours are very similar to stellar colours,
$\zem\sim2.5$--3.0, which is especially concerning for reddening
studies of DLAs. The magnitude selection criterion for most SDSS
quasar candidates ($i\le19.1$\,mag) is, by comparison, relatively
simple and, being applied in the $i$-band, will be less affected by
dust extinction than the bluer $u$, $g$ and $r$ bands. However, the
effect of dust on the $i$-band magnitude will increase as the absorber
redshift increases. Further, a fainter magnitude criterion of
$i<20.2$\,mag was applied to candidates which, from the photometric
colours, indicated the quasar was likely at high redshift,
$\zem>3.0$.

To address this concern and test the potential biases on our \EBV\
measurements, we have emulated the selection criteria for
spectroscopic follow-up of SDSS quasar candidates detailed by
\citet{Richards:2002:2945}. Our implementation and some basic
performance tests are described in \Aref{app:A}, and the code is made
publicly available in a portable form in \citet{Bernet:2015}. While we
hope this code has broad usefulness, we emphasise that it has been
tested mainly for its purpose in the current analysis; we encourage
other researchers to test it for their specific purposes. We also note
that there exist several additional, practical criteria that
determined whether a specific quasar candidate was followed-up
spectroscopically that are not reflected in our algorithm at all. For
example, ``fibre collisions'' between quasar candidates and
higher-priority targets for the SDSS (e.g.~low-$z$ galaxies) may mean
that a specific quasar candidate that passed the magnitude and
colour-selection criteria was not observed in reality. In this
respect, our emulation of the quasar selection algorithm should only
be used in a statistical way.

\begin{figure}
\begin{center}
\includegraphics[width=1.0\columnwidth]{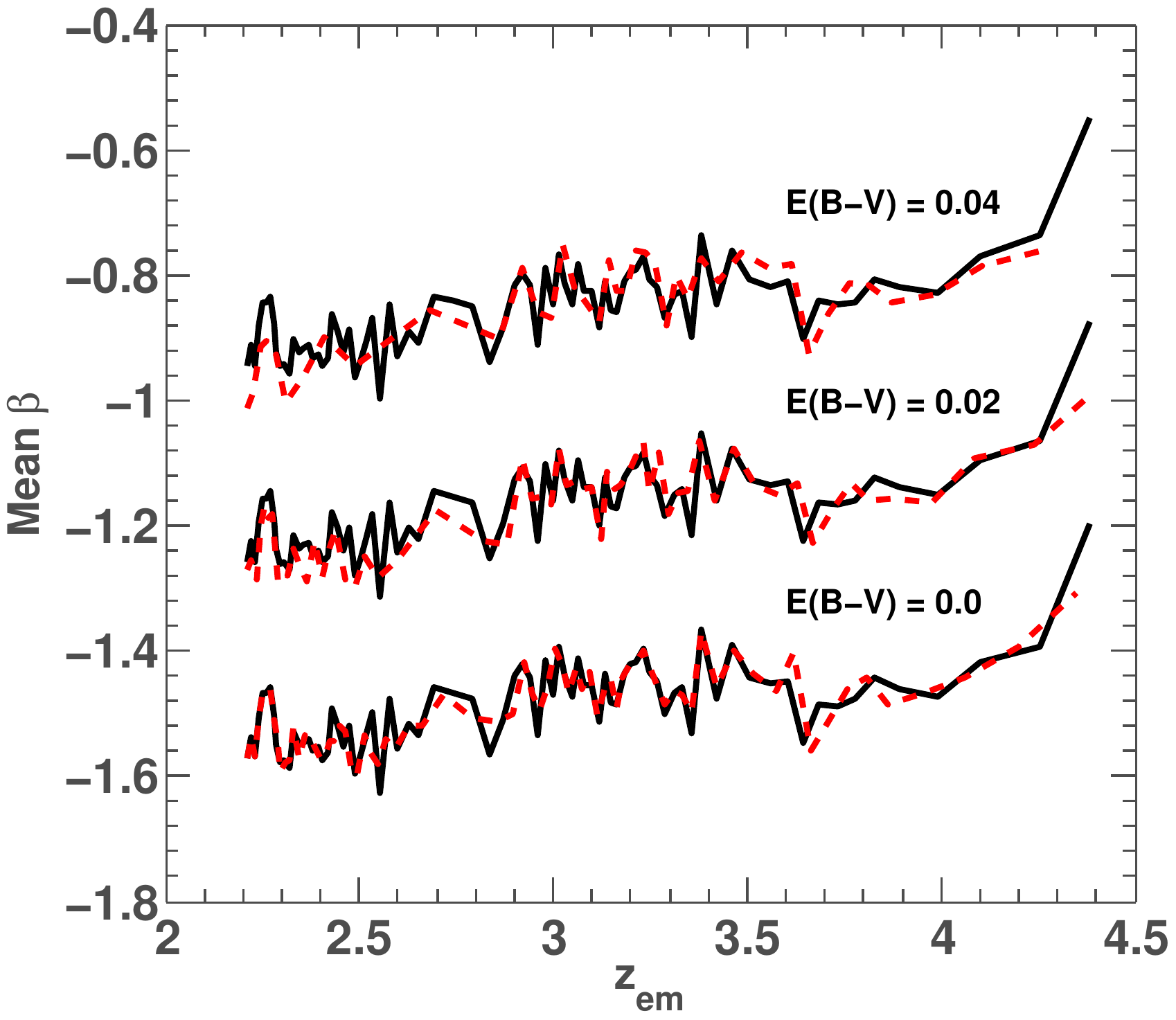}\vspace{-0.5em}
\caption{Mean spectral index, $\beta$, in bins containing 100 quasars,
  as a function of emission redshift, \zem, for the 7929 quasars in
  the \citet{Noterdaeme:2009:1087} sample falling in the relevant
  redshift range ($2.2\le\zem\le4.42$). The black solid lines show the
  mean $\beta$ for the original sample [i.e.~$\EBVSMC=0$\,mag] and
  where each quasar spectrum was reddened with $\EBVSMC=0.02$ and
  0.04\,mag. The red dashed lines show the mean $\beta$ of the reduced
  samples after passing the appropriately reddened photometry through
  our emulation of the SDSS quasar selection algorithm in
  \Aref{app:A}.}
\label{fig:beta_zem}
\end{center}
\end{figure}

With the quasar selection algorithm in hand, we can artificially
dust-redden each quasar spectrum, with a certain \EBV, and then
check whether they are selected by the quasar algorithm. This approach
enables the simple test shown in \Fref{fig:beta_zem}. Here we use the
entire sample of quasars in the \citet{Noterdaeme:2009:1087} catalogue
which fall in the relevant redshift range for our study,
$2.2\le\zem\le4.42$. This provides a 10-times larger sample than, say,
applying the test to just the DLA quasar sample alone. Three different
SMC-like reddening values [$\EBVSMC=0$, 0.02 and 0.04\,mag] are
applied, at the emission redshift, to each quasar's photometry and
spectrum, and the resulting $ugriz$ magnitudes are passed through our
selection code. \Fref{fig:beta_zem} compares the mean spectral index,
$\beta$ (determined as in \Sref{sec:analysis}), as a function of
emission redshift in bins containing 100 quasars, both before (black
line) and after (dashed red line) the selection code is applied.

\begin{figure}
\begin{center}
\includegraphics[width=1.0\columnwidth]{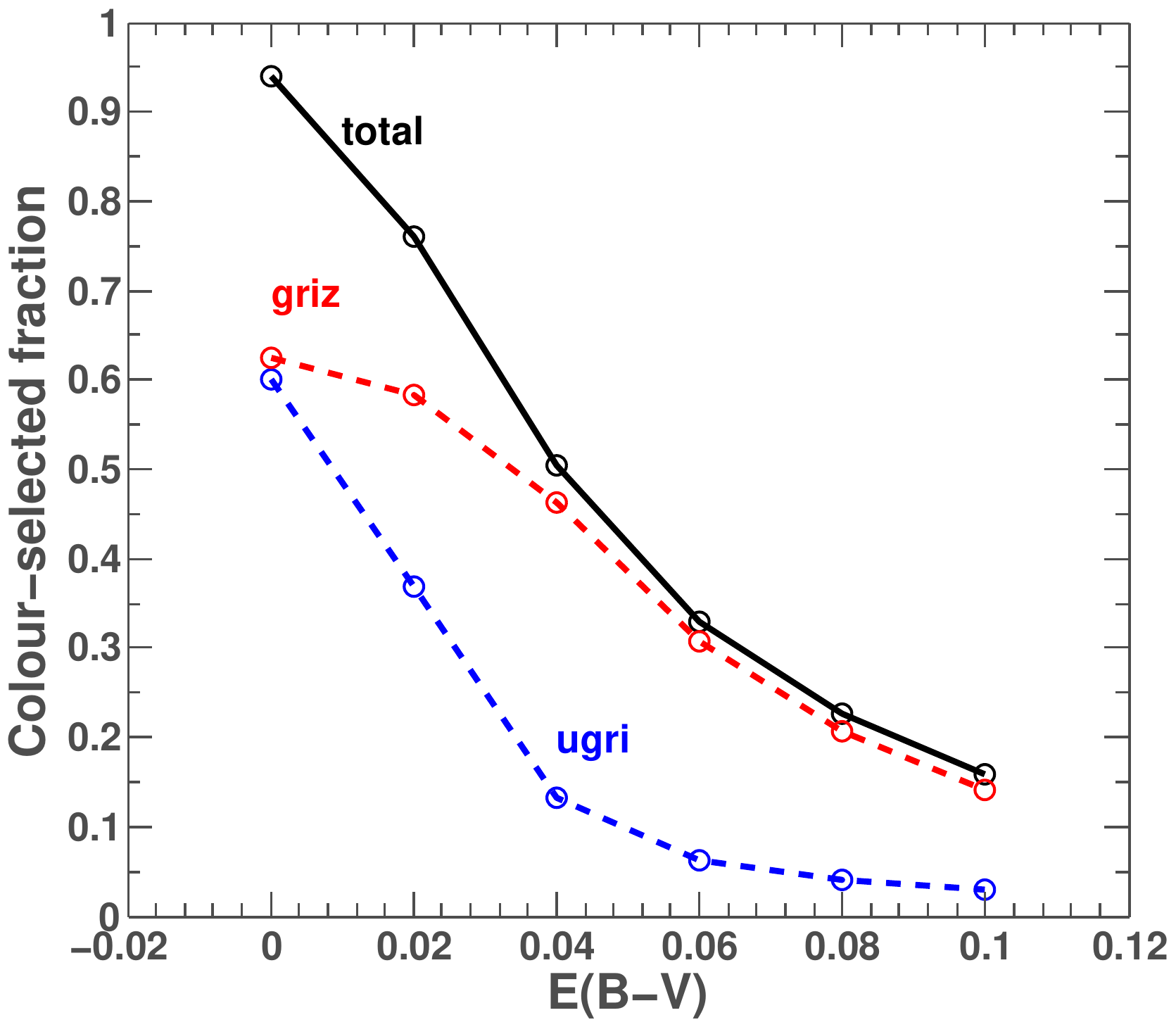}\vspace{-0.5em}
\caption{Fraction of quasars passing the magnitude and colour
  selection criteria (``colour-selected fraction'') in our emulated
  SDSS algorithm (\Aref{app:A}) as a function of the artificial
  reddening applied, \EBVSMC\ (at the quasar emission redshift), to
  the photometry. The blue (lower) and red (higher) dashed lines
  correspond to quasar selected via the $ugri$ (typically $\zem<3.0$)
  and $griz$ (typically $\zem>3.0$) channels, whereas the ``total''
  (black solid line) corresponds to those selected via either the
  $ugri$ or $griz$ channels. See \Aref{app:A} for further details.}
\label{fig:col-sel_frac}
\end{center}
\end{figure}

Firstly, \Fref{fig:beta_zem} demonstrates the simple, expected result
that reddening the spectra by an increment of $\EBVSMC=0.02$\,mag
increases the mean $\beta$ by $\sim$0.3 over the whole redshift
range. Secondly, focussing on the zero-reddening case, note that the
after-selection curve (red dashed line) does not completely match the
before-selection curve (black line). This is because not all quasars
in the \citet{Noterdaeme:2009:1087} catalogue are colour-selected;
$\approx$9\,per cent of the SDSS DR7 quasars were selected by other
criteria, such as radio loudness or X-ray emission
\citep{Schneider:2010:2360}. \Fref{fig:col-sel_frac} shows the
fraction of quasars that pass the magnitude and colour-selection
criteria in our code -- which we refer to as the ``colour-selected
fraction'' for brevity -- is a function of \EBVSMC. The
colour-selected fraction at zero reddening is similar to that
expected, and the fraction remaining at $\EBVSMC=0.003$\,mag -- the
mean value we find for SDSS DLAs -- is just $\approx$2\,per cent lower
than at zero reddening. Nevertheless, the fraction drops very quickly
with increased reddening, and just $\approx$52\,per cent remain if a
reddening of $\EBVSMC=0.04$\,mag is applied. See \Aref{app:A} for a
description of the channels through which a quasar can escape the
magnitude and colour-selection criteria.

Finally and most importantly, \Fref{fig:beta_zem} clearly demonstrates
that, despite the $\sim$50\,per cent decrease in the colour-selected
fraction at $\EBVSMC=0.04$\,mag, the mean $\beta$ is almost unaffected
and, at all redshifts, is very similar to the mean $\beta$ before the
magnitude and colour-selection criteria are applied. Indeed, the
difference between the before- and after-selection curves for the
$\EBVSMC=0.04$\,mag case is $<$5\,per cent of the absolute shift in mean
$\beta$ (i.e.~from the zero-reddening case). The largest effects are only
seen at the lowest and highest redshifts ($\zem\la2.4$ and $\ga$4.4),
so they will have little bearing on our DLA reddening results because
the fewest DLAs are found in quasars at these emission redshifts (see
\Fref{fig:redshift_dist}). We therefore conclude that the magnitude
and colour-selection criteria for spectroscopic follow-up of SDSS
quasar candidates has had a negligible effect on our reddening
estimates for (sub-)DLAs.

\section{Discussion}\label{sec:discussion}

\subsection{Comparison with recent DLA dust reddening measurements}\label{ssec:comparison}

\begin{figure}
\begin{center}
\includegraphics[width=1.0\columnwidth]{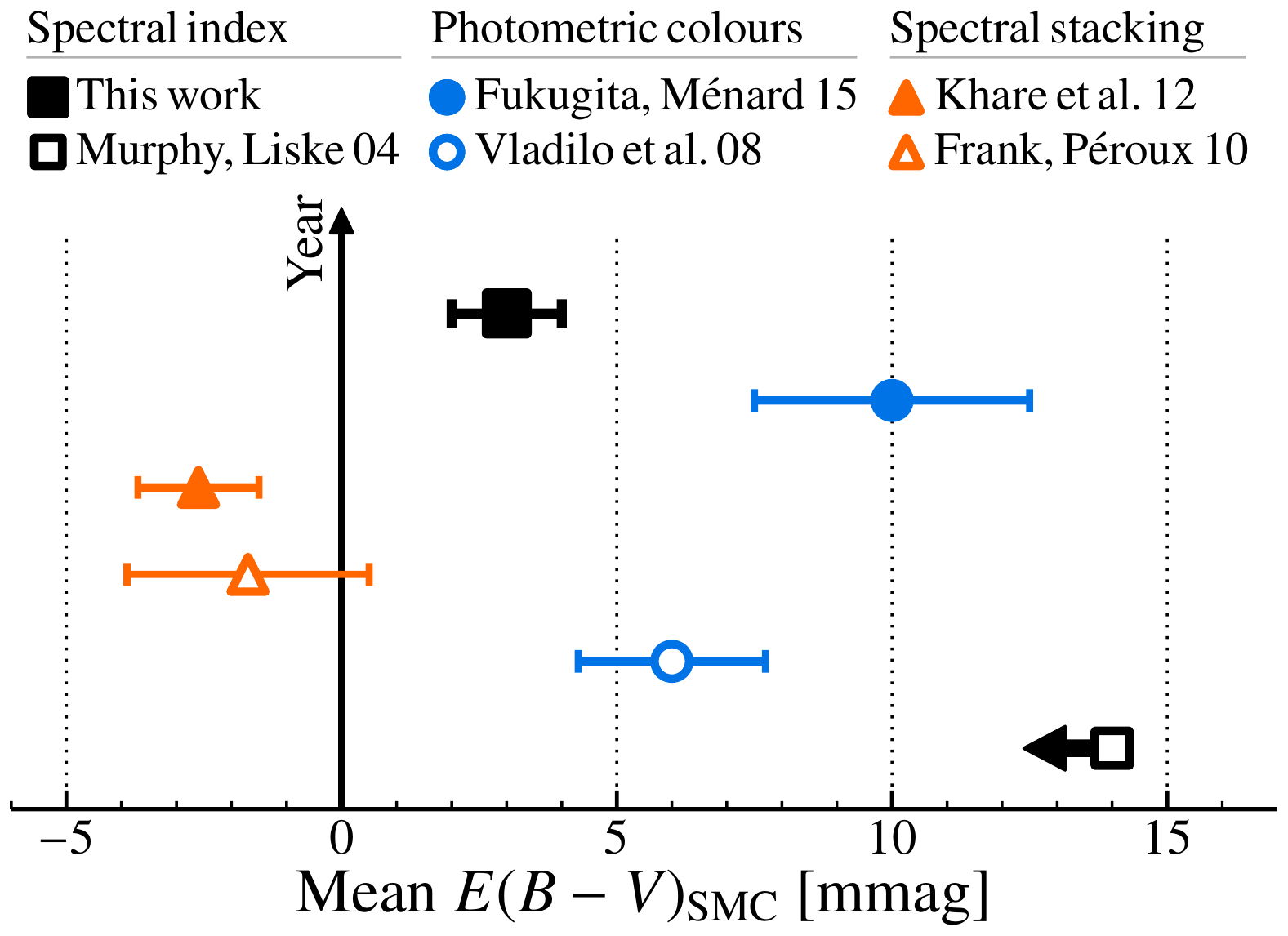}\vspace{-0.5em}
\caption{Summary of recent SDSS DLA dust reddening measurements,
  separated by methodology: analysis of spectral indices
  \citep[][and the present paper]{Murphy:2004:L31}, photometric
  colours \citep{Vladilo:2008:701,Fukugita:2015:195} and stacked
  spectra \citep{Frank:2010:2235,Khare:2012:1028}. The dust
  reddening is characterised by the mean colour excess for
  SMC-like dust, \EBVSMC, and 1-$\sigma$ error bars are given for
  all points (see text).}
\label{fig:EBVcomp}
\end{center}
\end{figure}

\Fref{fig:EBVcomp} compares our new DLA dust reddening measurement
with those from the other recent studies outlined in
\Sref{sec:introduction}. The six measurements shown are classified
according to the methodology used: quasar spectral index measurements,
as used in our work here and the early reddening limit from
\citet{Murphy:2004:L31}; the distribution of quasar photometric
colours, as analysed by \citet{Vladilo:2008:701} and, very recently,
\citet{Fukugita:2015:195}; or by analysing composite (`stacked') quasar
spectra, as in \citet{Frank:2010:2235} and \cite{Khare:2012:1028}. For
ease of comparison, all results are cast in terms of the mean colour
excess from SMC-like dust, \EBVSMC, over the whole sample
studied. Because all results derive from SDSS data, they all cover a
similar redshift range, $2.2\la\zab\la3.5$, except
\citet{Fukugita:2015:195} who restrict their DR9 $(g-i)$ colour analysis
to $2.1<\zab<2.3$ to avoid contamination of the $g$-band photometry
from \lya\ absorption. Finally, the measurements in \Fref{fig:EBVcomp}
are based on DLAs only, except that of \citet{Khare:2012:1028}: in
this case we report the \EBVSMC\ value from their `S1' comparison of
1084 quasars with a foreground $\lNHI\ge20.0$ absorber and the
appropriate control sample\footnote{For comparison with our results,
  the appropriate control sample is that which included quasars with no
  $\lNHI\ge20.0$ absorbers but which may include absorbers identified
  via metal-lines (at redshifts too low for detecting \Ion{H}{i}) --
  see table 1 of \citet{Khare:2012:1028}. The uncertainty in \EBVSMC\
  used in \Fref{fig:EBVcomp}, 1.1\,mmag, is the 1-$\sigma$ bootstrap
  error for their `S3' sample of 545 DLAs scaled by
  $\sqrt{545/1084}$.}.

\Fref{fig:EBVcomp} highlights the apparent disagreement between the
extant measurements of \EBVSMC. Our new measurement is consistent with
the photometric colour analysis of \citet{Vladilo:2008:701} at
$\approx$1.1-$\sigma$, but somewhat inconsistent with the higher
\EBVSMC\ value of $\approx$10\,mmag from the colour analysis of
\citet{Fukugita:2015:195} (i.e.~a $\approx$2.6-$\sigma$ difference). And
while those two previous colour analyses are consistent with each
other, they both differ significantly from the non-detection of
\citet{Frank:2010:2235} using spectral stacking (at $\approx$2.9 and
3.5-$\sigma$, respectively) and also the negative result from
\citet{Khare:2012:1028}. Our result differs from these two results
from spectral stacking as well. Clearly, taking all the existing
measurements at face-value, there is no clear indication which, if
any, indicate the correct mean dust reddening in SDSS DLAs. It is
therefore worth considering some differences between these
measurements, though it is beyond the scope of this paper to test in
detail whether they are responsible for the confusing picture in
\Fref{fig:EBVcomp} or not.

Firstly, all these recent results, except the photometric analysis by
\citet{Fukugita:2015:195}, rely on SDSS data subject to essentially
the same quasar colour selection algorithm. It is therefore unlikely
that the discrepancies between the results in \Fref{fig:EBVcomp} are
principally explained by such selection effects, especially given our
demonstration in \Sref{sec:bias} of their very weak influence. The DR9
quasar selection function differs substantially from earlier releases,
so it remains to be explored whether this might explain the higher
\EBVSMC\ value reported by \citet{Fukugita:2015:195}.

Secondly, \citet{Frank:2010:2235} pointed out some difficulties in
accurately measuring \EBVSMC\ from the ratio of stacked DLA and
non-DLA quasar spectra. They discuss how the emission line structure
of the quasar spectra bluewards of $\sim$1700\,\AA\ in the DLA
rest-frame, especially the well-known complexes of Fe lines, may limit
the accuracy of their ratio spectra in this region. Indeed, they
observe a distinct ``kink'' around $\sim$1700\,\AA\ in their ratio
spectra, noting that it was the main driver towards low \EBVSMC\
values. Visual inspection of their figures 4 \& 5 suggests a tendency
for the DLA--non-DLA ratio stack to be bluer at wavelengths
$\la$1800\,\AA\ and redder between
$\sim$1800--2500\,\AA. \citet{Frank:2010:2235} noted the composite
spectrum redwards of 1700\,\AA\ resulted in a more positive but still
null \EBVSMC. It is therefore possible that analysis of a restricted
wavelength range in stacking analyses may be consistent with our
result.

In our analysis, the control quasars for a given DLA quasar must
satisfy equations (\ref{eq:zcrit1}) and (\ref{eq:zcrit2}), i.e.~they
must lie at similar emission redshifts and their DLA search paths must
span the redshift of the DLA in question, respectively. The latter
criterion effectively imposes the same SNR requirements in the \lya\
forest of the control quasar spectra that allowed the detection of the
DLA in the DLA quasar's spectrum. That is, equations (\ref{eq:zcrit1})
and (\ref{eq:zcrit2}), together with the SDSS quasar magnitude and
colour selection criteria, ensure that the control quasars are
selected with the same restrictions that define the DLA quasar: the
same DLA could have been identified in all the control quasars at the
same redshift. \citet{Vladilo:2008:701}, \citet{Frank:2010:2235} and
\citet{Khare:2012:1028} did not impose \Eref{eq:zcrit2}, and instead
imposed a magnitude restriction: the control quasars needed to match
the corresponding DLA quasar's $i$ or $z$-band magnitudes ($\delta
z=0.5$\,mag, $\delta i=0.08$ and 0.4\,mag, respectively). However, not
imposing \Eref{eq:zcrit2} in our analysis would have artificially
biased our results towards zero or even negative reddening, as can be
seen via the following example. First, assume that DLAs cause no
reddening and that we did not impose equation (\ref{eq:zcrit2}) on
control quasars. The control quasars would therefore be allowed to
have lower SNR at bluer wavelengths -- i.e.~over the DLA search path
in the \lya\ forest -- than any corresponding DLA quasar could
have. This (together with the SDSS $i$-band magnitude selection
criteria) implies that the control quasars would have a redder
distribution of colours than the DLA quasars at any given
redshift. Restricting instead the $i$-band magnitude difference
between DLA quasars and their control quasars would not avoid this
bias. This is because control quasars would still be allowed to have
lower SNR at bluer wavelengths (over the DLA search path) and,
therefore, to have redder spectra, than any corresponding DLA quasar
could have, even if they all have the same $i$-band magnitude. Thus,
removing \Eref{eq:zcrit2} would tend to cancel any small reddening
signal from the DLAs in our analysis, or even result in a negative
mean \EBV, and imposing $i$-band magnitude matching would not alter
this bias substantially.

Finally, we note that \citet{Vladilo:2008:701} removed all identified
sub-DLAs and low-redshift metal-line absorbers from both their DLA and
non-DLA samples. Our approach, and that of most previous studies, is
to leave such contaminants in both samples, thereby avoiding the
significant task of identifying such absorbers. In principle, the
approaches are equivalent: these other absorbers are assumed to be
unrelated to the DLAs of interest, so should affect the DLA and
non-DLA/control samples in the same way and leave a negligible
residual in the differential measurement of \EBV.

As discussed in \Sref{ssec:EBV_metals}, one aspect of our new result
that lends additional confidence that the \EBV\ detection is really
due to DLA dust, is the observed correlation between \EBV\ and
metal-line strength (or metallicity) in Figs.~\ref{fig:EBV_EWSi},
\ref{fig:EBV_EWC} and \ref{fig:EBV_MSi}. \citet{Vladilo:2008:701} and
\citet{Khare:2012:1028} also found a marginal increase in reddening
for DLAs with $\EWSi>1$\,\AA\ ($\approx$1.5-$\sigma$ and
$\approx$2.1-$\sigma$, respectively), whereas \citet{Frank:2010:2235}
found no dependence on metal content. In our results, \EBV\ tends to
zero as the equivalent width of metal absorption approaches zero,
which further suggests that the \EBV\ values are accurate. A
systematic effect that could mimic this tendency is not immediately
obvious. Nevertheless, the lowest equivalent-width bins appear
slightly ($<$2-$\sigma$) below zero reddening, suggesting the
possibility that our \EBV\ measurements could be slightly
underestimated. The results of linear fits of \EBV\ vs.~\EWSi\ and
\EWC\ in \Tref{tab:res} provide weak evidence ($<$2-$\sigma$) that the
underestimate in \EBVSMC\ could even be $\approx$4\,mmag (cf.~the
overall average of 3\,mmag). A correction of this size would bring our
result into greater consistency with the photometric colour analyses
of \citet{Vladilo:2008:701} and \citet{Fukugita:2015:195}. However,
given the low (\textsl{post facto}) statistical significance of this
possibility, we do not make this correction here; testing it with
greater confidence awaits a significantly larger sample of DLA
quasars.

\subsection{The very low average dust content of SDSS DLAs}\label{ssec:lowdust}

It is clear from \Fref{fig:EBVcomp} that most DLAs towards SDSS
quasars contain very little dust. This was already apparent from the
limit imposed by the first SDSS study, \citet{Murphy:2004:L31}, which
was inconsistent with earlier results from small DLA and control
quasar samples, and it has clearly been confirmed by subsequent
studies. However, as noted above, \Fref{fig:EBVcomp} shows
considerable variance in the conclusions reached by different studies
about the mean DLA dust reddening.  There is not even a clustering of
values to provide a guide, only a hint that a reasonable mean value is
$\EBVSMC\sim2$--10\,mmag. Unfortunately, this observational status
precludes a detailed comparison of the mean dust content with other
probes of dust in DLAs, or with models for the effect on estimates of
\OgDLA. Nevertheless, some simple comparisons are illustrative.

One other clear signature of dust in DLAs is the relative depletion of
volatile and refractory elements (e.g.~Zn vs.~Fe, respectively) from
the gas phase onto dust grains. Such depletion indicators can be
determined in DLAs toward quasars bright enough to provide echelle or
echellette spectra with high SNR. For a ``typical'' DLA at
$\zab\sim3$, the metallicity may be ${\rm [Zn/H]}\approx-1.5$
\citep[e.g.][]{Rafelski:2012:89,Jorgenson:2013:482} and the dust
depletion characterised by ${\rm [Zn/Fe]}\approx0.3$
\citep[e.g.][]{Wolfe:2005:861}, implying $\EBVSMC\approx5$\,mmag for a
DLA with $\NHI=10^{21}$\,cm$^{-2}$ \citep{Murphy:2004:L31}. Clearly,
given the large ranges in DLA metallicity, column density and [Zn/Fe]
observed, \EBV\ is also expected to vary considerably, though a mean
$\EBVSMC>1$\,mmag appears likely. That is, a mean dust reddening of
$\EBVSMC\sim2$--10\,mmag is consistent with the dust depletion
signatures seen in DLAs towards brighter quasars. Naively, we might
have expected the latter to have systematically low dust content
because of the quasar brightness selection. However, it appears that
this must be a small effect. It is also plausible that it may be
overwhelmed by other effects, such as gravitational lensing, which
might bias such `echelle DLAs' to preferentially probe dustier
environments closer to galaxies than the mean SDSS DLA. A similar
effect may potentially explain why molecular hydrogen is detected
somewhat more often in `echelle DLAs' than in SDSS DLAs
\citep{Jorgenson:2014:2783}.

The extent to which dust in DLAs removes quasars, and therefore
high-\NHI\ DLAs, from optically-selected quasar samples, and the
possible consequences for \OgDLA, has remained an important concern
since it was raised by \citet{Ostriker:1984:1} and the measurements of
\citet{Pei:1991:6} implied it was indeed an important effect. However,
this concern has been substantially diminished by SDSS DLA reddening
studies and comparisons of DLA distributions in radio and optically
selected quasar samples
\citep[e.g.][]{Ellison:2001:393,Akerman:2005:499,Jorgenson:2006:730}.
Using a Bayesian analysis of a simple dust extinction model,
\citet{Pontzen:2009:557} considered the joint constraints offered by
these radio--optical comparisons and the photometric reddening results
of \citet{Vladilo:2008:701} to limit the likely fraction of DLAs
missing from optical samples to just 7\,per cent (1-$\sigma$
confidence). Replacing \citet{Vladilo:2008:701}'s estimate of the mean
SDSS DLA reddening with our new measurement, which is a factor of 2
smaller, would reduce this expected missing fraction below 5\,per cent
at 1-$\sigma$ confidence. And while it is always possible to
hypothesize that the dust extinction is bimodal in DLAs
\citep[e.g.][]{Khare:2007:487}, the dusty population is clearly small.
This seems consistent with the paucity of detections of the 2175\,\AA\
`dust bump' and 10\,$\umu$m silicate feature in DLAs
\citep{Junkkarinen:2004:658,Noterdaeme:2009:765,Kulkarni:2007:L81,Kulkarni:2011:14},
though the strong selection biases employed to find those systems make
a quantitative comparison difficult.

\section{Conclusions}\label{sec:conclusions}

Using spectral slope fits of the SDSS DR7 quasar spectra, and the
DLA/sub-DLA identifications of \citet{Noterdaeme:2009:1087}, we found
that the 774 selected quasars with a single foreground DLA are
significantly (3.2-$\sigma$) redder, on average, than carefully
selected control groups drawn from a sample of $\approx$7000 quasars
without foreground DLAs. The detection strengthens to 4.8-$\sigma$ if
sub-DLAs with $\lNHI\ge20.0$ are included (a total of 1069 absorber
quasars and $\approx$6700 control quasars). The reddening corresponds
to a mean, rest-frame colour excess due to SMC-like dust of
$\EBVSMC=3.0\pm1.0$\,mmag for DLAs at redshifts $\zab=2.1$--4.0
[$3.6\pm0.9$\,mmag when including sub-DLAs]. The DLA colour excess
correlates significantly with metal-line (rest-frame) equivalent
width, particularly \EWSi\ ($\approx$3.5-$\sigma$), with \EBVSMC\
increasing by $5.0\pm1.4$\,mmag per 1\,\AA\ increase in \EWSi, for
example. This provides further confidence in the reddening detection
and that it is really caused by dust in the DLAs. Weaker, less
significant correlations are seen with \Ion{C}{ii} equivalent width
and the metallicity derived from its known correlation with \EWSi.

No evolution of \EBV\ in DLAs with redshift was detected, with a
1-$\sigma$ limit of $\approx$2.5\,mmag per unit redshift for SMC-like
dust. However, this is consistent with the weak DLA metallicity
evolution observed in the same redshift range
\citep{Rafelski:2012:89,Jorgenson:2013:482} which, assuming a simple
proportionality with dust column, implies an increase in \EBVSMC\ of
only $\approx$1.3\,mmag per unit (decreasing) redshift. Similarly, no
significant dependence of \EBV\ on the neutral hydrogen column density
of DLAs is observed. Nevertheless, the data are consistent with a
linear relationship and the implied \EBV/\NHI\ ratio
[e.g.~$(3.5\pm1.0)\times10^{-24}$\,mag\,cm$^{2}$ for SMC-like dust] is
consistent with the ratios found in the Magellanic clouds after
accounting for the significantly lower typical metallicity of
DLAs. However, given the large spread in DLA metallicities, the
expected slope of any underlying dust--gas relationship in DLAs may
not be reliably comparable with that of the Magellanic clouds.

The very low dust-content of DLAs we find is consistent with that
implied by relative metal abundances derived from high-resolution
(echelle) spectra of typically much brighter, inhomogeneously selected
quasars. It is also consistent with the previously-reported
($\approx$3-$\sigma$) detection by \citet{Vladilo:2008:701} using SDSS
DR5 quasar photometric colours, but significantly smaller than that
implied by DR9 photometry \citep{Fukugita:2015:195} and inconsistent
with the null results from stacking DR7 quasar spectra
\citep{Frank:2010:2235,Khare:2012:1028}. Clarifying this somewhat
confusing observational picture of the mean DLA dust-content will
require further, independent measurements. Nevertheless, assuming
simple relations between dust extinction, \NHI\ and metallicity, and
including constraints from radio-selected DLA surveys, it remains
clear that DLAs dusty enough to be missing from optically-selected
samples are rare \citep[$\la$5\,per cent, cf.][]{Pontzen:2009:557}.

Finally, we demonstrated that the magnitude and colour selection of
SDSS quasars leads to $<$5\% bias in the reddening over the redshift
range of interest, $z=2.1$--4.0. Therefore, the above results do not
include corrections for, or systematic error components from, this
potential effect. Our code for emulating the SDSS colour-selection
algorithm of \citet{Richards:2002:2945} is publicly available in
\citet{Bernet:2015}. Significantly improving the precision with which
dust-reddening can be measured in DLAs will rely on substantially
increasing the sample of quasar spectra with accurate photometry
and/or spectrophotometry. Although much larger catalogues of quasars
are now available from SDSS \citep[e.g.~DR10,][]{Paris:2014:A54}, a
more accurate measurement of DLA reddening against the
($\approx$25-times larger) natural spread in quasar colours may demand
detailed accounting for the quasar selection criteria.

\section*{Acknowledgements}

We thank J.~Xavier Prochaska and Jochen Liske for many helpful
discussions and Pasquier Noterdaeme for providing the redshift paths
search for DLAs in his DR7 catalogue. We acknowledge the anonymous
referee for helpful comments that clarified important aspects of the
manuscript. We thank the Australian Research Council for a
\textsl{QEII Research Fellowship} (DP0877998) and for
\textit{Discovery Project} grant DP130100568 which supported this
work.

Funding for the SDSS and SDSS-II has been provided by the Alfred
P. Sloan Foundation, the Participating Institutions, the National
Science Foundation, the U.S. Department of Energy, the National
Aeronautics and Space Administration, the Japanese Monbukagakusho, the
Max Planck Society, and the Higher Education Funding Council for
England. The SDSS Web Site is
{\urlstyle{rm}\url{http://www.sdss.org/}}. The SDSS is managed by the
Astrophysical Research Consortium for the Participating
Institutions\footnote{The Participating Institutions are the American
  Museum of Natural History, Astrophysical Institute Potsdam,
  University of Basel, University of Cambridge, Case Western Reserve
  University, University of Chicago, Drexel University, Fermilab, the
  Institute for Advanced Study, the Japan Participation Group, Johns
  Hopkins University, the Joint Institute for Nuclear Astrophysics,
  the Kavli Institute for Particle Astrophysics and Cosmology, the
  Korean Scientist Group, the Chinese Academy of Sciences (LAMOST),
  Los Alamos National Laboratory, the Max-Planck-Institute for
  Astronomy (MPIA), the Max-Planck-Institute for Astrophysics (MPA),
  New Mexico State University, Ohio State University, University of
  Pittsburgh, University of Portsmouth, Princeton University, the
  United States Naval Observatory, and the University of Washington.}.







\appendix

\section{Emulating the SDSS quasar selection algorithm}\label{app:A}

\begin{figure}
\begin{center}
\includegraphics[width=1.0\columnwidth]{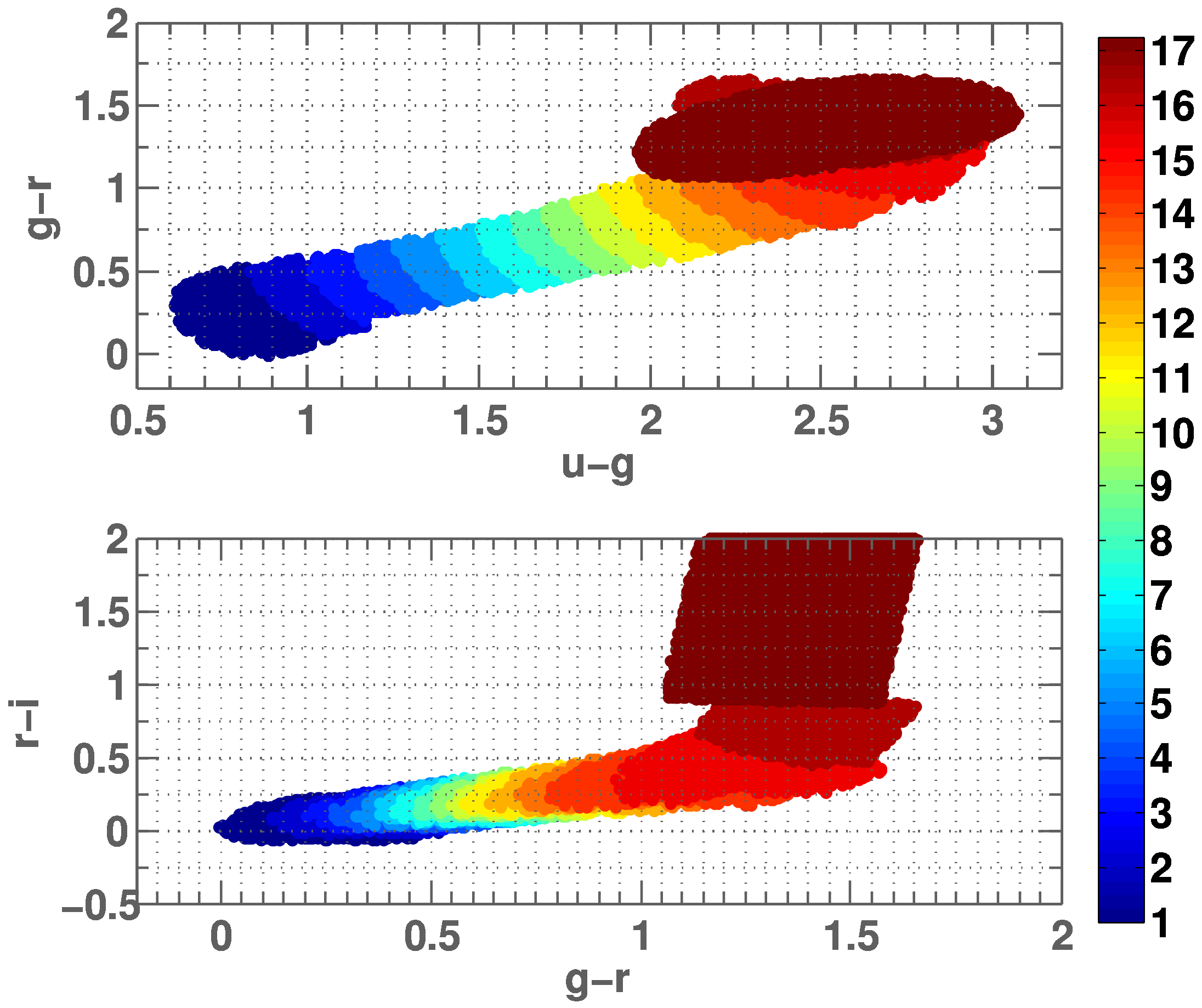}\vspace{-0.5em}
\caption{The $g-r$ versus $u-g$ and $r-i$ versus $g-r$ colour--colour
  planes of the stellar locus in our emulation of the ``low-z''
  (i.e.~$ugri$) SDSS quasar selection criteria
  \citet{Richards:2002:2945}. This figure is very similar to their
  figures 2 \& 3. The colour of a plotted point encodes the number of
  the nearest ``stellar locus point'', as defined in table 3 of
  \citet{Richards:2002:2945}. The plotted points define the projection
  of the 4-$\sigma$ error surface and were determined by Monte Carlo
  sampling. Objects with zero photometric errors will be selected as
  quasar candidates if they fall outside this surface. We note that,
  although our stellar loci plots are very similar to those ones of
  \citet{Richards:2002:2945}, there are some small differences that we
  can not account for, so users of our code should test it
  specifically for their purposes to ensure reliable statistical
  results.}
\label{fig:stl_ugri}
\end{center}
\end{figure}

\begin{figure}
\begin{center}
\includegraphics[width=1.0\columnwidth]{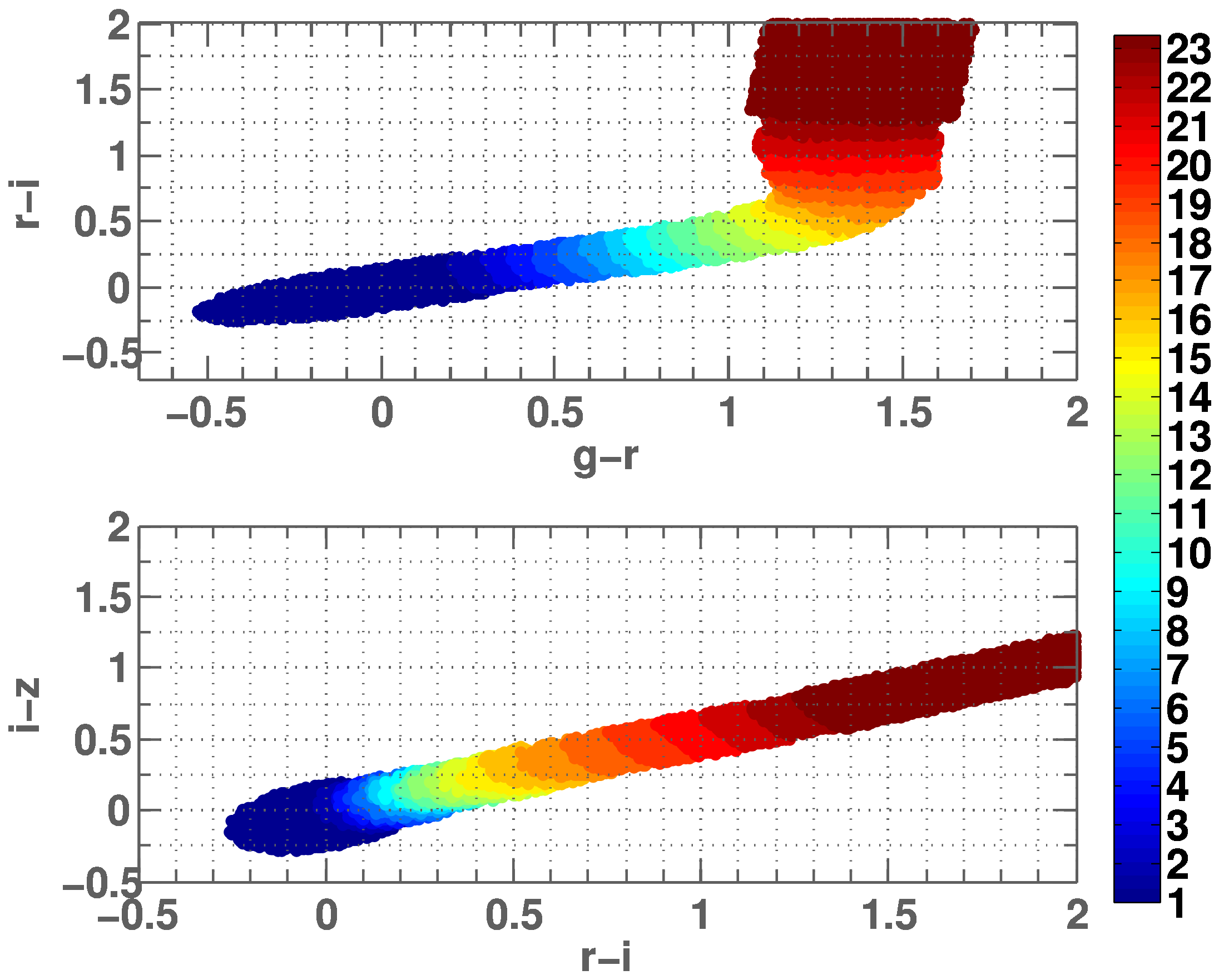}\vspace{-0.5em}
\caption{Same as \Fref{fig:stl_griz} but for the $r-i$ versus $g-r$
  and $i-z$ versus $r-i$ colour--colour plane for the ``high-z''
  (i.e.~$griz$) quasar selection criteria. This figure is very similar
  to figures 5 \& 6 of \citet{Richards:2002:2945} and the colour of a
  plotted point encodes the number of the nearest ``stellar locus
  point'' defined their table 4.}
\label{fig:stl_griz}
\end{center}
\end{figure}

In this Appendix we briefly review the approach followed by
\citet{Richards:2002:2945} to select quasar candidates, based on their
SDSS photometric magnitudes and colours, for spectroscopic
follow-up. We the demonstrate that our emulation of this approach
performs appropriately for the purposes of this paper, primarily via
the results in \Fref{fig:col-sel_frac} and Figs.~\ref{fig:stl_ugri} \&
\ref{fig:stl_griz}.

SDSS quasar candidates were selected via their colours in $ugriz$
broadband photometry, specifically the target point-spread-function
(PSF) magnitudes in these filters. Broadly speaking, objects were
selected as quasar candidates if they were consistent with being point
sources and if they fell outside the four dimensional colour space
inhabited by the stellar locus \citep{Richards:2002:2945}. The colours
of ordinary stars occupy a continuous, narrow region in the `colour
space' ($u-g$)--($g-r$)--($r-i$)--($i-z$) \citep{Newberg:1997:89} and
the stellar temperature is the main parameter that determines the
position along this stellar locus. And while stars have approximately
blackbody spectra, quasars spectra can be characterized by a power-law
overlaid with broad emission lines. Therefore, quasars generally have
colours quite distinct from stellar colours and can be identified as
outliers from the stellar locus. The exception to this is the
well-known redshift range, $2.5\la\zem\la3.0$ where quasar colours
cross the stellar locus.

\citet{Newberg:1997:89} modelled the stellar locus as a
two-dimensional ribbon with a varying elliptical cross section; more
specifically, their locus consisted of a series of overlapping right
elliptical cylinders capped with half-ellipsoids. To simplify the
algorithm for quasar colour-selection, \citeauthor{Richards:2002:2945}
split the four-dimensional colour space into two three-dimensional
colour spaces, $ugri$ and $griz$. In essence, an object is then
considered a quasar candidate if its colours lie outside the ellipse
given by the convolution of the closest stellar colour ellipse and the
ellipse formed from the object's photometric colour
errors.

The SDSS quasar selection algorithm consists of checking an object's
target PSF magnitudes and uncertainties against four main criteria:
\begin{enumerate}
\item[1.] Whether they lie in an `exclusion box', i.e.~regions of the
  colour space in which the too-numerous Galactic contaminants lie,
  such as white dwarfs, white dwarf pairs, A stars and M stars.
\item[2.] Whether they lie more than 4-$\sigma$ outside the $ugri$
  stellar locus and $i\le19.1$ (corresponding mainly to quasars at
  $\zem<3.0$).
\item[3.] Whether they lie more than 4-$\sigma$ outside the $griz$
  stellar locus and $i\le20.2$ (corresponding mainly to quasars at
  $\zem>3.0$).
\item[4.] Whether they lie in an `inclusion region' of colour space
  \citep[see the specific definitions in section 3.5.2
  of][]{Richards:2002:2945}. These apply to objects whose colours are
  consistent with being quasars at:\vspace{-0.8em}
  \begin{enumerate}
  \item[4.i] $2.5<\zem<3.0$, 10\,per cent are selected (based on
    their right ascension) from those near, but still more than
    2-$\sigma$ outside, the stellar locus (i.e.~closer than the usual
    4-$\sigma$ exclusion criterion);
  \item[4.ii] $\zem\le2.2$, a simple $u-g$ colour selection is applied;
  \item[4.iii] $\zem>3.6$, which are not well selected by the $griz$
    outlier criteria, further $griz$ selection criteria are defined;
  \item[4.iv] $\zem>4.5$, a selection using $riz$ colours is defined.
  \end{enumerate}
\end{enumerate}
If an object satisfied criteria 1 and 2, or 1 and 4.i, or 1 and
4.ii, \citeauthor{Richards:2002:2945} classified it as a ``low-$z$
quasar''. If it object satisfied criteria 1 and 3, or 1 and 4.iii, or
1 and 4.iv, it was classified as a ``high-$z$ quasar''. One object may be
classified as both a low-$z$ and high-$z$ quasar in this way.

We implemented the stellar locus definition and these criteria into a
{\sc matlab} code to emulate the SDSS quasar selection algorithm of
\citeauthor{Richards:2002:2945}, which we provide in
\citet{Bernet:2015}. Potential users of this code should note that
several channels for selecting quasar candidates in
\citeauthor{Richards:2002:2945} were not implemented in our code. For
example, candidates that are radio sources in the FIRST survey
\citep{Becker:1995:559} were generally included by
\citeauthor{Richards:2002:2945} (see their figure 1). This makes their
selection independent of colour, but for this reason it was not
important to implement this in our code to test the colour-selection
sensitivity of our reddening
analysis. \citeauthor{Richards:2002:2945}'s algorithm also deals with
objects that were not detected in some filters. However, this is only
important for very faint quasar candidates, or those at $\zem\ga4.7$
(where the $g$-band flux may be entirely consumed by a Lyman-limit
system). These cases are not relevant for our sample selection, so we
neglect this complication in our code. We also do not implement the
criteria used to distinguish point sources from extended
objects. Finally, note that the \citet{Schneider:2010:2360} DR7 quasar
catalogue also contains quasars which were selected as part of a
galaxy and luminous red galaxy survey, as showing X-ray emission, in
different stellar surveys and via serendipitous detection, as
specified in their table 1.

Figures \ref{fig:stl_ugri} and \ref{fig:stl_griz} show the relevant
colour--colour planes of the 4-$\sigma$ error surfaces for the $ugri$
and $griz$ stellar loci in our code. Objects with zero photometric
errors will be selected outside this surface. In practice, points
plotted in these diagrams were determined by a simple Monte Carlo
test: artificial objects covering the $ugriz$ space, all with
$\sigma_{u,g,r,i,z}=0.0$, were tested against criteria 2 and 3 above
and those objects not selected as quasar candidates were
plotted. Figures \ref{fig:stl_ugri} and \ref{fig:stl_griz} are very similar
to figures 2 \& 3, and 5 \& 6 in \citeauthor{Richards:2002:2945}
and demonstrate that the stellar locus is adequately modelled in our
code. Nevertheless, some small differences are visually apparent
(despite several checks to ensure our code matches the written
specifications) and, although these should not have any important
effects on our dust-reddening analysis here, we caution that other
researchers should test our code for their specific purposes to ensure
the reliability of any conclusions they draw.

A further, simple test of our code is to check whether it selects
quasar candidates via the same selection channel as specified by
\citet{Schneider:2010:2360}. For this test we use the entire sample of
$2.2<\zem<4.42$ quasars in the \citet{Noterdaeme:2009:1087} catalogue
from which our reddening analysis stems. Inputting the target PSF
photometry of these quasars into our code results in 60\,per cent of
them being selected via the $ugri$ channel and 62\,per cent via
the $griz$ channel (noting that each object can be selected by either,
both or neither channel). These values are similar to, though slightly
higher than, the values of 51 and 59\,per cent, as defined by the
classification of \citet{Schneider:2010:2360} using the same selection
algorithm as in \citet{Richards:2002:2945}. Again, although these
differences are not important for assessing the degree of bias
entering our dust-reddening analysis (especially in light of the
results in \Fref{fig:beta_zem}), it is possible that such differences
are important for other high-precision, statistical studies with large
samples, so we again recommend further, specific reliability tests in
those circumstances.


\section*{Supporting Information}\label{sec:supp}

Additional Supporting Information may be found in the online version
of this article:\vspace{-0.5em}\newline

\noindent \textbf{\Tref{tab:DLA_sample}.} The DLA quasar sample (774
  quasars).\vspace{-0.1em}\newline
\noindent \textbf{\Tref{tab:Abs_sample}.} The absorber quasar sample (1069
  quasars).\vspace{-0.1em}\newline
\noindent \textbf{\Tref{tab:Non-DLA_sample}.} The non-DLA quasar sample (7105
  quasars).\vspace{-0.1em}\newline
\noindent \textbf{\Tref{tab:Non-Abs_sample}.} The non-absorber quasar sample (6733
  quasars).\vspace{-0.5em}\newline

\noindent Please note: Oxford University Press are not responsible for the
content or functionality of any supporting materials supplied by
the authors. Any queries (other than missing material) should be
directed to the corresponding author for the paper.

\bsp	
\label{lastpage}
\end{document}